%% file: IGW.tex
\documentclass[fleqn,usenatbib]{mnras}
\newcommand{\figwidth}{\columnwidth}
 
\usepackage{newtxtext,newtxmath}
\usepackage{hyperref}
\hypersetup{
    colorlinks=true,
    linkcolor=blue,
    filecolor=magenta,      
    urlcolor=cyan,
}
\usepackage[T1]{fontenc}

\DeclareRobustCommand{\VAN}[3]{#2}
\let\VANthebibliography\thebibliography
\def\thebibliography{\DeclareRobustCommand{\VAN}[3]{##3}\VANthebibliography}

\usepackage{graphicx}	
\usepackage{amsmath}	

\usepackage{ulem}

\input{paper_resources/macros/formatting}

\input{paper_resources/macros/symbols}
\input{paper_resources/macros/units}  
\input{paper_resources/macros/vectors}

\usepackage{soul}
\newcommand{\brunt}{Brunt--V\"ais\"al\"a}
\newcommand{\papI}{Paper I}
\newcommand{\papIII}{Paper~III}	
	
\newcommand{\microhz}{\mathrm{\umu{} Hz}}
\newcommand{\code}[1]{\texttt{#1}}
\newcommand{\mesa}{\code{MESA}}
\newcommand{\GYRE}{\code{GYRE}}
\newcommand{\ppmstar}{\code{PPMstar}}
\newcommand{\los}{line-of-sight}
\newcommand{\komega}{$l-\nu$ diagram}
\newcommand{\npeak}[0]{$N^2$-peak}

\title[Convection and waves in massive main-sequence stars II]{ 3D
  hydrodynamic simulations of massive main-sequence stars II.
  Convective excitation and spectra of internal gravity waves}

\author[W. Thompson et al.]{William Thompson$^{1,2}$,
Falk Herwig$^{2,\dagger}$\thanks{E-mail: fherwig@uvic.ca},
Paul R. Woodward$^{3,\dagger}$,
Huaqing Mao$^{3,\dagger}$,
\newauthor{}
Pavel Denissenkov$^{2,\dagger}$,
Dominic M. Bowman$^{4,5}$,
Simon Blouin$^{2,\dagger}$\\
$^{1}$NRC Herzberg Astronomy and Astrophysics, Victoria B.C., V9E 2E7, Canada\\
$^{2}$Department of Physics \& Astronomy, University of Victoria, Victoria, B.C., V8W 2Y2, Canada\\
$^{3}$LCSE and Department of Physics and Astronomy, University of Minnesota, Minneapolis, MN 55455, USA\\
$^{4}$School of Mathematics, Statistics and Physics, Newcastle University, Newcastle upon Tyne, NE1 7RU, United Kingdom\\
$^{5}$Institute of Astronomy, KU Leuven, Celestijnenlaan 200D, B-3001 Leuven, Belgium\\
$^{\dagger}$Joint Institute for Nuclear Astrophysics - Center for the Evolution of the Elements (JINA-CEE)
}

\date{Accepted: 26 April 2024. Submitted: 3 November 2022.}

\pubyear{2024}

\begin{document}
\label{firstpage}
\pagerange{\pageref{firstpage}--\pageref{lastpage}}
\maketitle
\normalem

\begin{abstract}
Recent
photometric observations of massive stars have identified a
low-frequency power excess which appears as stochastic low-frequency variability in light curve observations. We present the oscillation
properties of high resolution hydrodynamic simulations of a \unit{25}{\Msun}
star performed with the \ppmstar~code. The model star has a convective
core mass of $\approx \unit{12}{\Msun}$ and approximately half of the
envelope simulated.
From this simulation, we extract light curves
from several directions, average them over each hemisphere, and
process them as if they were real photometric observations.
We show how core convection excites waves with a similar frequency
as the convective time scale in addition to significant power across a forest of low and high angular degree $l$ modes.
We find that the coherence of these modes is relatively low as a result of their
stochastic excitation by core convection, with lifetimes
on the order of 10s of days.
Thanks to the still significant power at higher $l$ and this relatively low
coherence, we find that integrating over a hemisphere produces a
power spectrum that still contains measurable power up to the Brunt--V\"ais\"al\"a frequency.
These power spectra extracted from the stable envelope are
qualitatively similar to observations, with same order of magnitude yet lower characteristic frequency.
This work further shows the
potential of long-duration, high-resolution hydrodynamic simulations
for connecting asteroseismic observations to the structure and dynamics of core
convection and the convective boundary.

\end{abstract}


\section{Introduction}
\lSect{intro} Asteroseismology has evolved into a powerful tool in
stellar physics, to probe stellar interiors and to validate our
basic understanding of stellar physics and evolution. For example,
asteroseismology has been able to constrain the extent and nature of
convective boundary mixing in main-sequence stars
\citep[e.g.][]{Aerts:2003fz,Moravveji:2015kr,Moravveji:2016jq,Angelou:2020ih,Viani:2020hr,michielsen:21,Bowman2021c}
and the theoretical studies suggest that even more detailed
constraints may be possible in the future
\citep{Pedersen:2018ew,Michielsen:2019ht,Pedersen2021a}.

When a star's hydrostatic equilibrium is perturbed, different kinds of
oscillations can result. Familiar sound waves arise from pressure
perturbations and propagate at the local speed of sound. On longer
time scales, a second class of oscillation can also occur when there is
a density disturbance. In this case, the force of gravity acts to
counteract the over- or under-density, so they are referred to as
gravity or buoyancy waves \citep{ASTERO_BOOK}.  Gravity waves can occur in stratified
fluids either at an interface between two disparate fluids, in which
case they are called surface (or interfacial) waves, or in a continuously stratified
fluid, in which case they are called internal gravity waves (IGWs).
IGWs obey a dispersion relation where their wave vectors make lie at angles $\varphi$
with respect to surfaces of a constant density according
to $\omega=N \cos(\varphi)$, where $N$ is the
Brunt--V\"ais\"al\"a (BV) frequency. This quantity sets the maximum
frequency at which gravity waves can propagate.

Gravity waves are important probes of stellar interiors, as they are
sensitive to the physics of stellar structure and are particularly
sensitive to the interior mixing and rotation profiles within massive
stars \citep{Rogers:2013fl,Aerts:2015jv,Bowman2020c}.  In
asteroseismology, standing gravity waves are referred to as g modes,
whereas standing pressure waves are referred to as p modes.
We follow the convention that p modes are labelled with positive radial order $n$ and g modes are labelled with negative radial order $n$ (higher radial order g modes are more negative).

In massive stars, gravity waves are excited at the interface of
convective and radiative regions such as the boundary of the
hydrogen-burning convective core during the main-sequence phase of
stellar evolution and partial ionization zones within the radiative envelope \citep{Dziembowski1993e,Dziembowski1993f}. Since convection is a turbulent and intermittent process, the
excitation is stochastic. These gravity waves have been
hypothesized as significant contributors to photometric variability in
massive stars, especially at low frequencies where coherent
heat-driven g modes excited by local opacity enhancements in the
iron-bump layer are not the dominant source of variability \citep{Aerts:2015jv,Bowman:2019ka}.

Recently,
high-precision photometric observations of hundreds of massive stars
assembled by the NASA K2 and TESS missions \citep{Howell2014,
  Ricker2015} have revealed the distinct features of stochastic low-frequency
variability inferred to be caused by gravity waves
\citep{Bowman:2019ka}. From such observations,
\citet{Bowman2020b,Bowman2022e} demonstrated
that the morphology of the stochastic low-frequency variability (i.e. dominant
periodicities and amplitudes) directly probes the evolutionary
properties of the host star (i.e. its mass and age).

From a stellar structure model using a given set of input physics,
such as those available from the 1D evolution code
\mesa\ \citep{Paxton2011, Paxton2013, Paxton2015, Paxton2018,
  Paxton2019}, pulsation codes such as \GYRE\ are able to calculate the
resonant eigenmodes for various pulsation mode geometries of a stellar
model \citep{townsendGYREOpensourceStellar2013,
  townsendAngularMomentumTransport2018}. Such calculations for a grid
of models allow forward seismic modelling of stars based on a
quantitative comparison of the observed and theoretically-predicted
pulsation mode frequencies
\citep{Aerts2021a}. Furthermore, non-adiabatic
calculations reveal the growth rate of pulsation modes and allow the
excitation physics of heat (i.e.\ opacity) driven coherent pulsations
to be probed. However, as demonstrated by recent space-based
observations of massive stars, there remain large uncertainties
associated with rotation and stellar opacities which significantly
affect the predictions of opacity-driven coherent mode excitation
(e.g. \citealt{Burssens:2020bz}).

To study the excitation of gravity waves by convection,
astrophysicists have turned to two and three dimensional hydrodynamic
simulations \citep{Ratnasingam:20,Horst:2020ds,lesaux:2023,Ratnasingam:2023,anders:2023}. These codes directly simulate the motions and properties
of fluids.
\citet{Rogers:2013fl} and
\citet{edelmannThreeDimensionalSimulationsMassive2019} have surmised
that the excitation of stochastic IGWs in massive stars is driven
predominantly by convective plumes that overshoot the convective
boundary between the core and the stable envelope. 
Some of these simulations
required very high heating factors beyond the nominal luminosity of
the star (on the order of $10^6$ times higher) and correspondingly
high viscosities to maintain numerical stability. 
These simulations showed
plumes tend to excite g modes with large wavelengths, or low angular
degrees $l$.  

In \papI\ \citep{Herwig:2023} we presented 3D stellar
hydrodynamic simulations with heating factors ranging from $10^{1.5}$ to $10^4$ and high grid resolution based on the explicit, compressible gas-dynamics approach using the \ppmstar\ code. Those simulations allow us to characterize the large-scale and turbulent nature of core convection in massive stars and establish the converged excitation spectrum up to angular spherical harmonic degree $l \approx 100$. 

The global flow morphology leads to boundary layer separation flows in
which small-scale instabilities are generated causing
perturbations. This is reflected in the flat spectrum of the
radial component of convective motions near the boundary compared to a fully-developed
Kolmogorov spectrum in the core region far away from the boundary.

The stable layers immediately outside the convective core host IGWs with the dominant power in the radial velocity component at radial order $n=-1$ and large
wave number $l \approx 70$.
These IGWs have been hypothesized as a source of mixing in the
stable layers adjacent to the convective boundary \citep[][and discussion there]{Blouin:22a}. 
The IGW spectrum of the horizontal velocity component follows the turbulent spectrum maintaining the familiar $\propto l^{-5/3}$ power law spectrum, whereas the radial velocity component assumes a distinctly different spatial spectrum that peaks at high wave numbers $l \approx 30 \dots 70$. Reynolds stresses cause a tight correlation between the horizontal spectrum in the convection and stable layers, but a broad spectrum of radial plume-like excitations facilitate the distinct spectral morphology of the radial velocity component.

In this paper we  explore the properties of IGWs further out in the stable layers and present a detailed theoretical asteroseismic analysis of three
simulations with heating factor $1000\times$. The  goal of this
paper is to expand our analysis of IGWs from the boundary region explored in \papI\ to the stable envelope further away from the boundary, and to establish the spectral properties of synthetic observations of light curves of the oscillating envelope. 
We are motivated by the observed asteroseismic features of massive stars such as the near-ubiquitous low-frequency variability reported by \citet{Bowman:2019ka}, but note that our simulations do not include the outer envelope and surface that are probed by real observations, nor the effects of radiative damping.
Despite these limitations, this work nonetheless aims to identify mechanisms and effects that could play a role in the origin of a spectrum with a low-frequency excess. 
 
We begin in \Sect{methods} by
describing our methods including a brief account of the underlying
hydrodynamic simulations, the asteroseismic analysis technique, and the \code{GYRE} analysis of the spherically averaged radial profiles for the mode identification.
In \Sect{results}, we present the flow morphology of convection in the core and oscillations in the envelope. Then, we present simulated synthetic light curves extracted as if we were observing the luminosity of the star partway through the radiative envelope. We perform an asteroseismic analysis of these light curves and reveal a strong low-frequency excess. From there, the remainder of the paper is dedicated to dissecting this luminosity power spectrum and explaining how its various features come to be. Finally, \Sect{discussion} presents discussion and conclusions.

\section{Methods}\lSect{methods}
    
Based on 3D hydrodynamic simulations of a \unit{25}{\Msun} main-sequence
star slightly evolved beyond the zero-age main sequence \citep[\papI,][]{Herwig:2023} we generate simulated
observations of light curves to compare them with observations. We
decompose the oscillations in three dimensions into spherical
harmonics, and compare them with eigenmodes calculated using GYRE to
identify specific IGW modes.

\subsection{Hydrodynamic simulations}\lSect{hydro-sim}
\begin{table}
  \centering
  \caption{The most important simulations from \papI\ analyzed in this
    paper. Additional simulations featured only in \Sect{s.convergence}
    are described in detail in \papI.
     The table provides the run name, the grid size,
    number of dumps (see details on \ppmstar\ dumps in
    \papI), total length of run in simulated star time.} \lTab{tab:run_table}
  \begin{tabular}{lrrr}
      \hline 
      Name & Resolution & Dumps & Duration \\
      \hline
      M107 &  $ 768^3$ & 9943 & \unit{290}{\days} \\
      M114 & $1152^3$ & 5926 & \unit{170}{\days} \\
      M115 &  $1728^3$ & 3584 & \unit{103}{\days} \\
  \hline 
  \end{tabular}
\end{table}
The simulations used here are summarized in \Tab{tab:run_table} and
described in more detail in \papI. The simulations have been performed
with the \ppmstar\ gas dynamics code
\citep{woodwardHydrodynamicSimulationsEntrainment2015,
  jonesIdealizedHydrodynamicSimulations2017,andrassy3DHydrodynamicSimulations2020,
  stephens3D1DHydronucleosynthesisSimulations2021},
with several important updates.  The most relevant for this work
relates to \ppmstar\ now solving the conservation laws in terms of
perturbations with respect to a base state which aids computational
accuracy. It also makes the mapping of a 1D stratification from a
stellar evolution code to 3D substantially easier and results in an initial 3D
stratification that is by definition in hydrostatic equilibrium. At the bounding sphere we impose a reflecting boundary condition using ghost cells that mirror the cells across the bounding surfaces (see section 2.2 in \papI\ for details).

In our $25\Msun$ simulation, the convective boundary is located at $\approx 1500 \Mm$ on top of which is a stable envelope with a radial extent of $1200\Mm$. The \mesa\ model has a radius of
$5000\Mm$ and  therefore the simulation includes $54\%$ of the star's radius.
Since we
dedicate a significant fraction of our resolution budget to the core,
the simulations resolve the large-scale flows and turbulence, as well as interactions between convective flows and the boundary. This allows the simulation to accurately capture the excitation process. We include enough of the stable envelope to probe oscillation properties several pressure scale
heights away from the convective boundary without getting close to the
outer edge of the simulation domain.

The base state of the 3D hydrodynamic simulations approximates the
structure of a  \unit{25}{\Msun} stellar evolution model calculated with the
\mesa\ stellar evolution code \citep[][\texttt{template} run]{Davis2019}. 
Details are provided in \papI. The model star is
 near the zero-age main-sequence, $\natlog{1.64}{6}\yr$ after the
start of H burning, and its central H mass fraction has decreased to
$X(\mathrm{H})_\mathrm{c} = 0.606$ from the initial $0.706$.

The three simulations differ by grid resolution. In all cases the
output (dump) cadence is $\approx 43\minute$, which is comparable to
the observational cadence of the 30-min (``long-cadence'') observing mode
of the K2 and TESS\footnote{High value TESS
targets are recorded with much higher cadence.} planet-hunting satellites \citep{Howell2014,
  Ricker2015}.
The $768^3$ simulation M107 was
followed for slightly over nine months of simulated time. The long
baseline and correspondingly high resolution for low frequencies allow
at least qualitative comparisons to asteroseismic observations. However, with
regard to some mixing properties this grid resolution is not entirely
sufficient.  Simulation M114 with a $1152^3$ grid still follows
\unit{170}{\days} of simulated time and is ideal to study the
internal structure of the star and its oscillations. The M115 simulation was
performed on a $1728^3$ grid for \unit{103}{\days} containing over
five billion grid cells and is used to determine the convergence
properties of the oscillation results. 

\begin{figure}
    \centering
    \includegraphics[width=\figwidth]{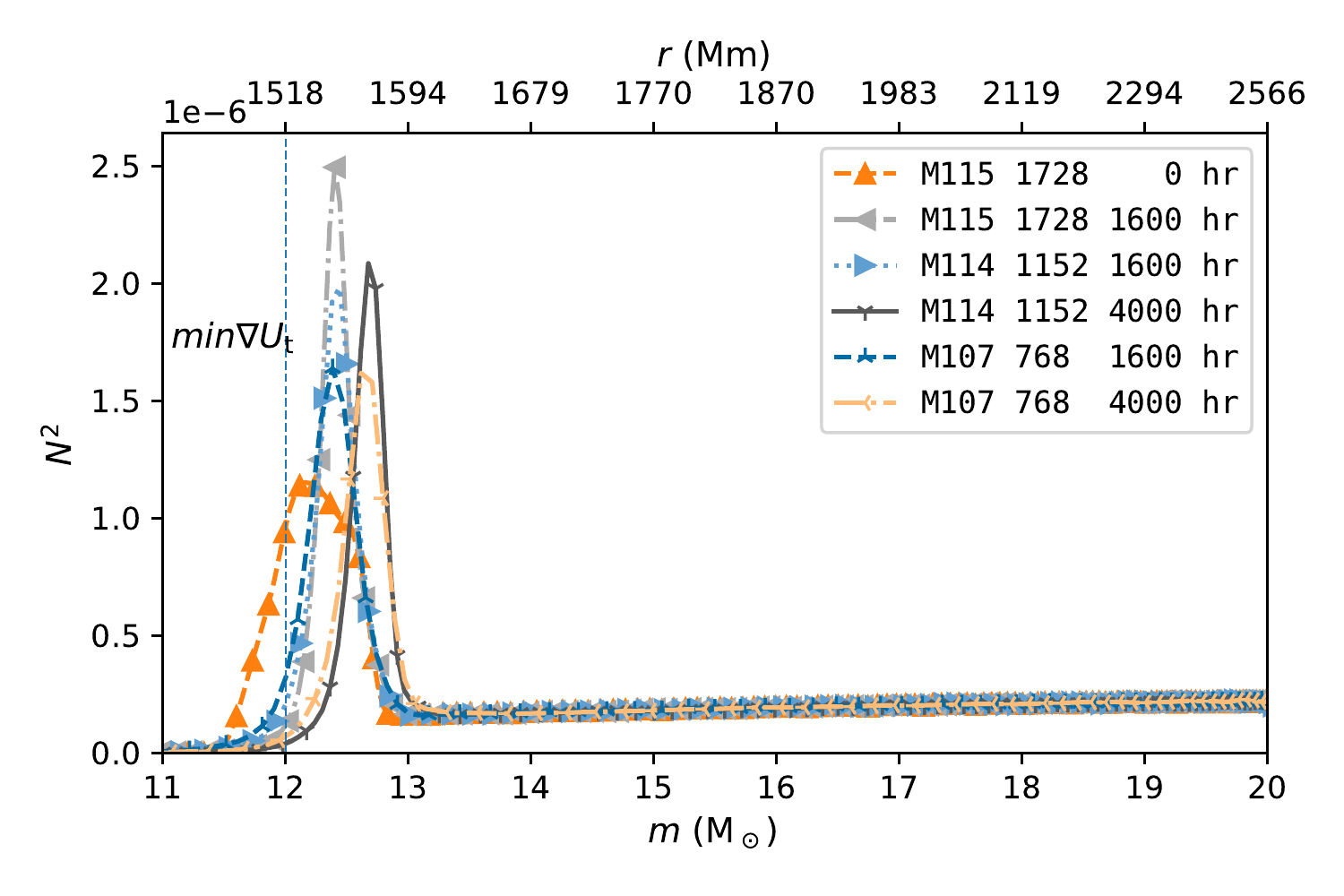}
    \caption{Profile of BV frequency for the three simulations. $N^2$
      has been calculated from the spherically averaged profiles of
      the 3D hydrodynamic simulations. The labels indicate the run ID,
      the grid resolution and the time of the simulation for
      which each profile is shown. The vertical dashed line indicates
      the location of the minimum gradient of the tangential velocity
      component for run M107, $t=1600\hour$ which is discussed in
      \papI\ as one criterion to define the convective
      boundary. $N$ has units of the angular frequency $\mathrm{rad\ s^{-1}}$.}
      \lFig{brunt}
\end{figure}
The \brunt\ frequency profile of the adopted base state and its changes during
subsequent simulation run-time are shown in \Fig{brunt}. Despite applying a
heating factor of $1000$ which makes the convective flows $10\times$
faster compared to nominal heating, the flows remain very slow. It
takes about \unit{40}{d} of simulated time until a steady state is
reached. 
By then, it has established a convective boundary profile due to hydrodynamic processes only. This can be seen in the evolution of the $N^2$ peak profile in \Fig{brunt} (cf.\ Fig.\ 15 in \papI). At higher resolutions, the maximum of the \npeak\ is larger and the boundary region is narrower. In the real star the shape of the \npeak\ in the boundary is the convolution of its dynamic-, thermal-, and nuclear-time scale evolution.
The base states we adopted for these simulations reflect the convective boundary mixing assumptions adopted in the MESA stellar evolution model and may have to be revised in future.
Note that the values in \Tab{tab:run_table} are the total simulated time, including the initial transition period.

Fourier analysis of the oscillations requires a long duration baseline, but we nonetheless
avoid the initial phase in which the $N^2$ boundary is
changing because it might affect details of the wave excitation. We
therefore exclude the initial $1300$ dumps ($930\hour$) from the
analysis. 

As the simulations run, the following data are output for later analysis. Rendered images of the simulation are created from full resolution two-dimensional outputs at each dump (\Fig{movie_panels}; Section 2.2 of \papI\ for
details). Likewise, one dimensional radial profile outputs prepared by the simulation code are based
on uncompressed, full resolution data at the time of each dump. This includes asteroseismic observables (\Fig{lums}). Analysis of the three dimensional structure of the simulation, on the other hand, is carried out against reduced resolution data 
\citep[that we call \emph{briquette} data, see ][for details]{stephens3D1DHydronucleosynthesisSimulations2021} that contains three dimensional outputs of some quantities with a resolution reduced by a factor 4 in each direction but full 4-byte precision. The decomposition of oscillation motions into spherical harmonics is based on this type of filtered output data.

\subsection{Asteroseismic analysis}\lSect{astero-analysis}

The \ppmstar\ code generates simulated observations of the
luminosity fluctuations by first calculating the bolometric radiance $L$ of each
octant of each grid cell using its temperature and treating it as a
black body. These simulated observations are generated at each radius in the simulation up to $54\%$ of the radius of the star. The luminosity is instead reported at mass
coordinates up to $m_\mathrm{r} = \unit{20}{\Msun}$.  

Eight different vantage points from which
the simulated star might be observed are chosen not to align with the simulation
grid.  The first four lines of sight (los) to the star are chosen to
be along:
\begin{eqnarray*}
\overrightarrow{\mathrm{los}_1}=&(1,1,1)\\
\overrightarrow{\mathrm{los}_2}=&\overrightarrow{\mathrm{los}_1}\times(0,0,1)\\
\overrightarrow{\mathrm{los}_3}=&\overrightarrow{\mathrm{los}_1}\times\overrightarrow{\mathrm{los}_2}\\
\overrightarrow{\mathrm{los}_4}=&\overrightarrow{\mathrm{los}_1}+\overrightarrow{\mathrm{los}_2}+\overrightarrow{\mathrm{los}_3}
\end{eqnarray*}
Each vector is normalized before it is used to define the others.
The subsequent four are anti-parallel to the first four,
i.e. $\overrightarrow{\mathrm{los}_5}=-\overrightarrow{\mathrm{los}_1}$
and
$\overrightarrow{\mathrm{los}_6}=-\overrightarrow{\mathrm{los}_2}$,
etc. This choice allows us to examine the simulation from all sides
and avoids grid-alignment.  The radiance $L$ is integrated over one
hemisphere for each \los\  and for each radius. We apply
Lambert's factor $\cos(\theta)$ where $\theta$ is the angle
between the surface normal and the \los.  This factor changes
the surface integral into an integral over the projected face of the
star, mimicking an observation of an unresolved point source (though of course, we do not simulate the surface and the boundary condition is different between the simulation and a real star).

This surface integral is evaluated by computing the volume integral of $L
\cos(\theta)$ in one hemisphere of a thin shell divided by its
thickness at each radius. This is repeated at every dump of the
simulation, giving a time sequence of eight simulated photometric
observations at every radius. In \Sect{sec:attenuation}
we demonstrate the impact of applying Lambert's cosine factor which
deemphasizes the outer annulus of the integration, and thereby the
contribution of modes with high angular degree $l$. We also
demonstrate how approximating the full sphere integral with just one
point on the hemisphere
\citep[see for example ][]{edelmannThreeDimensionalSimulationsMassive2019} overemphasizes
the power at frequencies just below the BV frequency.

In long duration simulations there is a slow expansion of the
core. To account for this we work in Lagrangian coordinates converting
our radial luminosity profiles into mass coordinates at each time
step.  For each simulation run, we analyze the time series luminosity
data as if they were real observations of a star. As noted in \papI,
these simulations neglect the effects of radiation. As heat is added
in the core, the luminosity throughout the star slowly increases. We
begin by removing this trend to the luminosity. 

\begin{figure}
    \includegraphics[width=\columnwidth]{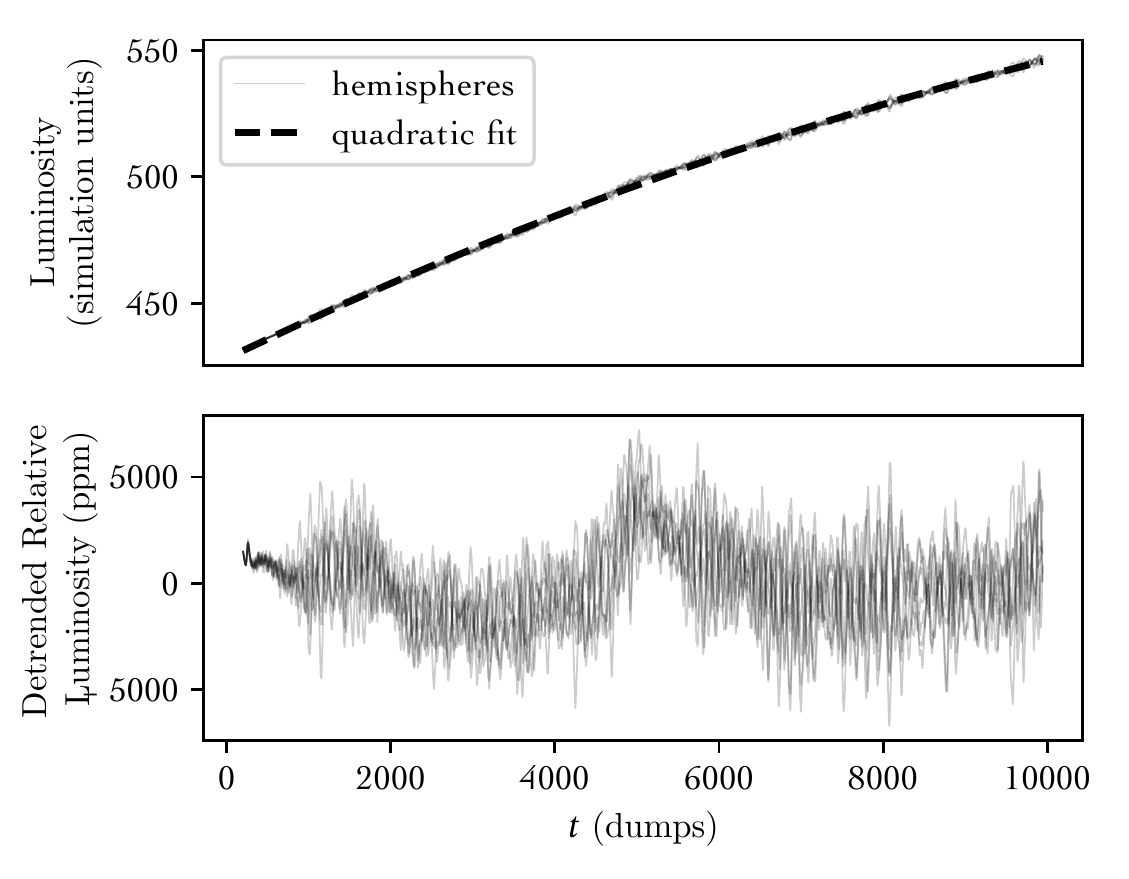}
    \caption{We take out the global luminosity trend (top panel) by fitting a
      quadratic to the average of all eight lines of sight. Dividing
      the luminosity time series from each \los\ by this fit
      gives the relative luminosity.  The time axis is in units of \emph{dumps} which are written out with a frequency of $\frac{1}{\unit{43}{\minute}}$. The second panel shows the unity-subtracted fluctuations in luminosity compared to the quadratic fit.}  \lFig{detrend}
\end{figure}
\Fig{detrend} shows how we detrend the integrated luminosity
curves at each mass coordinate by fitting a quadratic, and dividing
through to arrive at the relative luminosity. We then apply a Hanning
window function to impose a periodic boundary condition and control
spectral leakage, and take discrete Fourier transforms of each time
series to arrive at power spectra (\Fig{lums}). 
We correct the amplitude spectra for power
lost due to the window function.
In the simulations listed in \Tab{tab:run_table} the default dump frequency is approximately \unit{370}{\microhz}.

These light curves and their resulting spectra may appear different since they have none of the typical noise sources present in real observations such as read noise or photon noise. The only noise source present is numerical noise, which as we will see in \Sect{s.convergence}, does not significantly effect our study.
As a result, certain subtle effects like spectral leakage may be visible.
For this same reason, we believe it is valid to interpret signals close to the Nyquist limit and to the minimum resolvable frequency (besides the very lowest frequencies that we have removed with our luminosity detrending).

To investigate differences and similarities to observed spectra (\Sect{low-freq-excess}
) we re-produce the procedure applied to real time-series
photometric observations of massive stars
(e.g. \citealt{Bowman:2019ka}), we apply the
method known as iterative pre-whitening
\citep{Degroote2009a,Papics2012a,Bowman_BOOK}.
As an iterative procedure, we identify the frequency and
amplitude of the highest-amplitude and statistically significant peak in the amplitude spectrum of
the light curve, optimise the frequency, amplitude and phase using a
non-linear least-squares fit to the light curve using
        \begin{equation}
  \Delta\,m = A\,\cos\left(2\pi\nu\left(t - t_0\right) + \phi\right)
          \end{equation}
where $A$ is the amplitude, $\nu$ is the frequency, $\phi$
is the phase, $t$ is the time with respect to a zero-point $t_0$. The
optimised sinusoid is then subtracted from the light curve to produce
a residual light curve. From the residual light curve, a residual
amplitude spectrum is calculated and the next iteration proceeds. In
the study of coherent pulsation modes in intermediate- and high-mass
stars, it is typical to continue iterative pre-whitening until an
amplitude signal-to-noise (S/N) criterion is satisfied. Here we use
the common value of an amplitude S/N $\geq 4$ to determine if a peak
is significant, in which the noise is calculated using a \unit{1^{-1}}{d}
window centred at the frequency value of extracted peak at each
iteration. In this way, our iterative pre-whitening procedure only
excludes the few dominant high-amplitude peaks in the amplitude spectrum,
which lie on top of the low-frequency power excess background
\citep{Bowman:2019ib}.

To follow the same methodology as applied to observations, we fit the model
\begin{equation}
\alpha(\nu) = \frac{\alpha_0}{1 + \left( \frac{\nu}{\nu_{\rm
      char}} \right)^{\gamma}} + C_{\rm w} ,
\end{equation}    
where $\alpha_0$ is the amplitude at zero frequency, $\gamma$ is the
logarithmic amplitude gradient, $\nu_{\rm char}$ is the characteristic
frequency, and $C_{\rm w}$ is a frequency-independent (i.e. white)
noise term \citep{Bowman:2019ib} using a least-squares regression fit to
the logarithm of the residual (i.e. post iterative pre-whitening)
simulation data. This model thus yields $\nu_{\rm{char}}$ and $\gamma$
that can be directly compared to their observed
counterparts. 

To produce spectrograms, also known as sliding Fourier transforms, we
divide the time series into sub-sequences 512 dumps in length that
advance one dump at a time.
We apply a Hanning window to each sub-sequence and take
the Fourier transform to arrive at low resolution amplitude spectra. These
amplitude spectra are then stacked according to the time at the middle of
the sub-sequence to reveal how modes change over time. To reveal
pertinent features, we will show the power relative to a smoothed background
spectrum.
   
\subsection{Wavenumber-frequency decomposition}\lSect{k-omega-meth}
We generate wavenumber-frequency $l-\nu$ diagrams 
from the briquette data
\citep{stephens3D1DHydronucleosynthesisSimulations2021} to decompose
oscillations into both spatial and temporal frequencies as a
post-processing step. Since we are working with the modes of a
spherical object, we choose to compute the power using a spherical
harmonic basis (i.e. with discrete spherical
harmonic degree $l$ instead of continuous, replacing wavenumber $k$).
For each mass coordinate of interest, we sample
points across a regular grid of $\theta$ and $\phi$ angular
coordinates using a trilinear interpolation.\footnote{This analysis
uses methods implemented in the
\href{https://github.com/PPMstar/PyPPM}{PyPPM} library.}
Although the longest runs include almost $10\,000$ dumps corresponding to $>54$ convective turnover times, we use only $2000$ dumps for the frequency wavenumber analysis. This is because the convective boundary is migrating (cf.\ \Sect{stoch-excit} and \papI) which may cause artefacts if a longer baseline is used. In our spectra the frequency resolution is therefore \unit{0.185}{\microhz}.
                
We perform the spherical harmonic decomposition using the
\code{SHTools} library \citep{wieczorekSHToolsToolsWorking2018}. We
decompose the radial component of the velocity field first into
complex spherical harmonic coefficients for each time step.
Next, we apply a Hanning window along the time axis to
control spectral leakage.
Finally, we take the discrete Fourier transform of each time series of spherical harmonic coefficients.
To calculate the relative luminosity fluctuations, we divide the spherical harmonic coefficients
by the time-averaged $l=0$, $m=0$ coefficient for each radius.
No windowing function is required along the
spatial axes since the boundary conditions are naturally
periodic. Finally, we use the \code{SHTools} library to calculate the
power spectral density of the radial velocity oscillations normalized
by degree $l$ for each frequency bin. The spherical harmonic transform
and Fourier transform are both linear, so we can apply them in either
order. For high-resolution runs such as M115, this post-processing
step requires parallelization on a computer cluster.

The power by frequency, degree $l$, and mass coordinate represent a
reduction from four dimensions to three.
The resulting data cube can be sliced by mass to
give frequency-wavenumber diagrams at each radius, or by $l$ to
identify specific modes.

\subsection{Eigenmodes of the spherically-averaged radial stratification}\lSect{gyre-meth}
\lSect{s.eigenmodes}
We use the stellar oscillation code
\code{GYRE}\footnote{\url{https://gyre.readthedocs.io}}
\citep{townsendAngularMomentumTransport2018,townsendGYREOpensourceStellar2013}
to identify the eigenmodes of waves based on the spherically averaged
radial stratification of the 3D simulations. In this way we can
identify modes and features in the \komega s extracted from the 3D
simulations as described in \Sect{k-omega-meth}.  \GYRE\ solves a
two-point boundary value problem for the sets of the standard
equations of stellar oscillations for the adiabatic and non-adiabatic
cases with an option to include rotation. We consider the case of
linear adiabatic non-radial oscillations for which the governing
differential equations are summarized in Appendix A of
\cite{townsendGYREOpensourceStellar2013}. The code uses a novel Magnus
Multiple Shooting numerical scheme to find eigenfrequencies of
oscillations and their corresponding eigenfunctions representing
radial displacement and Eulerian perturbations of pressure and
gravitational potential. \GYRE\ is an open source Fortran code that
uses OpenMP directives for parallel computations on multiple cores. It
offers several integrators of differential equations among which we
have chosen the fourth-order Mono-Implicit Runge-Kutta method as the
most stable for our model.

We transform state variable radial profiles from the hydrodynamic
simulation dumps to the \mesa\ format readable by \GYRE.  Because the
hydrodynamic model has an insufficient number of radial zones
for the \GYRE\ code to work well, we map it into a larger number of
radial zones of its corresponding \mesa\ model using the Akima spline
interpolation. For the outer boundary condition in our GYRE calculations we used the vacuum condition with a zero pressure at the surface.

We run the \GYRE\ code for the harmonic degree $l$ ranging from 1 to
45 to identifity the resonant g mode oscillations of radial orders $-20 \leq n \leq -1$, f modes, and low-radial order p modes. The selected ranges of $l$
and $n$ numbers are determined by the successfulness of our
\GYRE\ computations.  From these computations we get the
eigenfrequencies $\nu$ of oscillation modes and their
corresponding radial $\xi_r$ and horizontal $L\xi_\mathrm{h}$
displacement amplitudes for a given dump of our hydrodynamic
simulations, where $L = \sqrt{l(l+1)}$.

\section{Results}\lSect{results}

\begin{figure*}
  \centering
  \includegraphics[width=\textwidth]{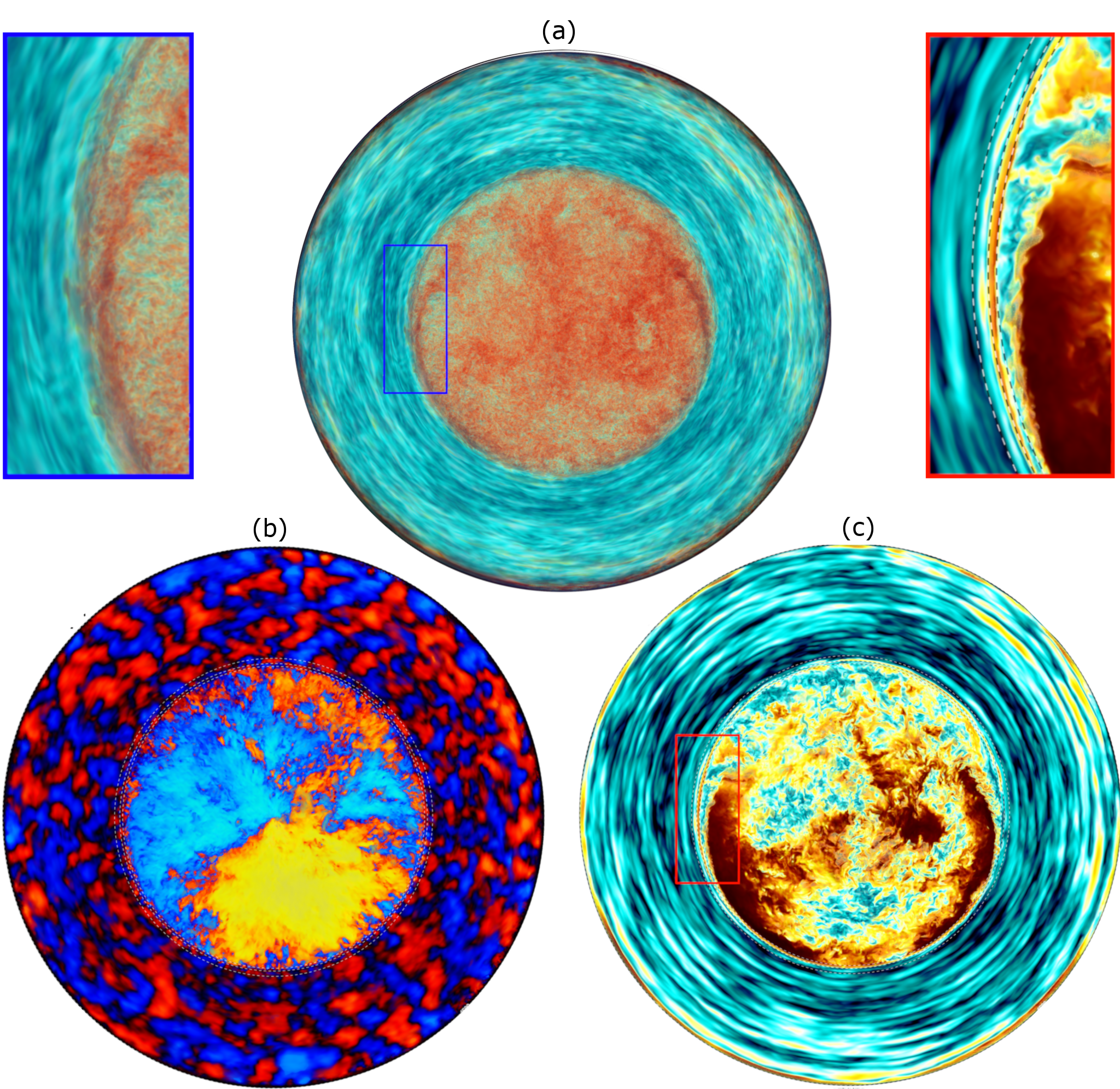}
  \caption{Centre-plane slice renderings from dump 1500 of run M115. \textbf{(a)} Vorticity. The color scheme has vorticity magnitude decreasing red, yellow, light and
    dark blue.
    \textbf{(b)} Radial velocity. The color scheme has radial velocity light to dark blue
    represent decreasing negative velocity magnitudes that are
    directed inward, yellow-orange-red represent decreasing positive
    radial velocity components.
    \textbf{(c)} Tangential velocity magnitude. The color scheme has tangential velocity 
    magnitude decreasing from dark brown, red, yellow, white, light and dark blue.
    The white-black dashed circles
    have been added at the positions of $\min \grad |U_\mathrm{t}|$
    (see Section 4.3 in \papI), the radius $r_\mathrm{N^2-peak}$
    where $N^2$ has a maximum and $r_\mathrm{N^2-peak} + 0.15\Hpzero$. Full resolution images and movies for runs M114 and M115 are available at 
    \href{https://www.ppmstar.org}{https://www.ppmstar.org.}}
  \lFig{movie_panels}
\end{figure*}

\begin{figure*}
\includegraphics[width=\textwidth]{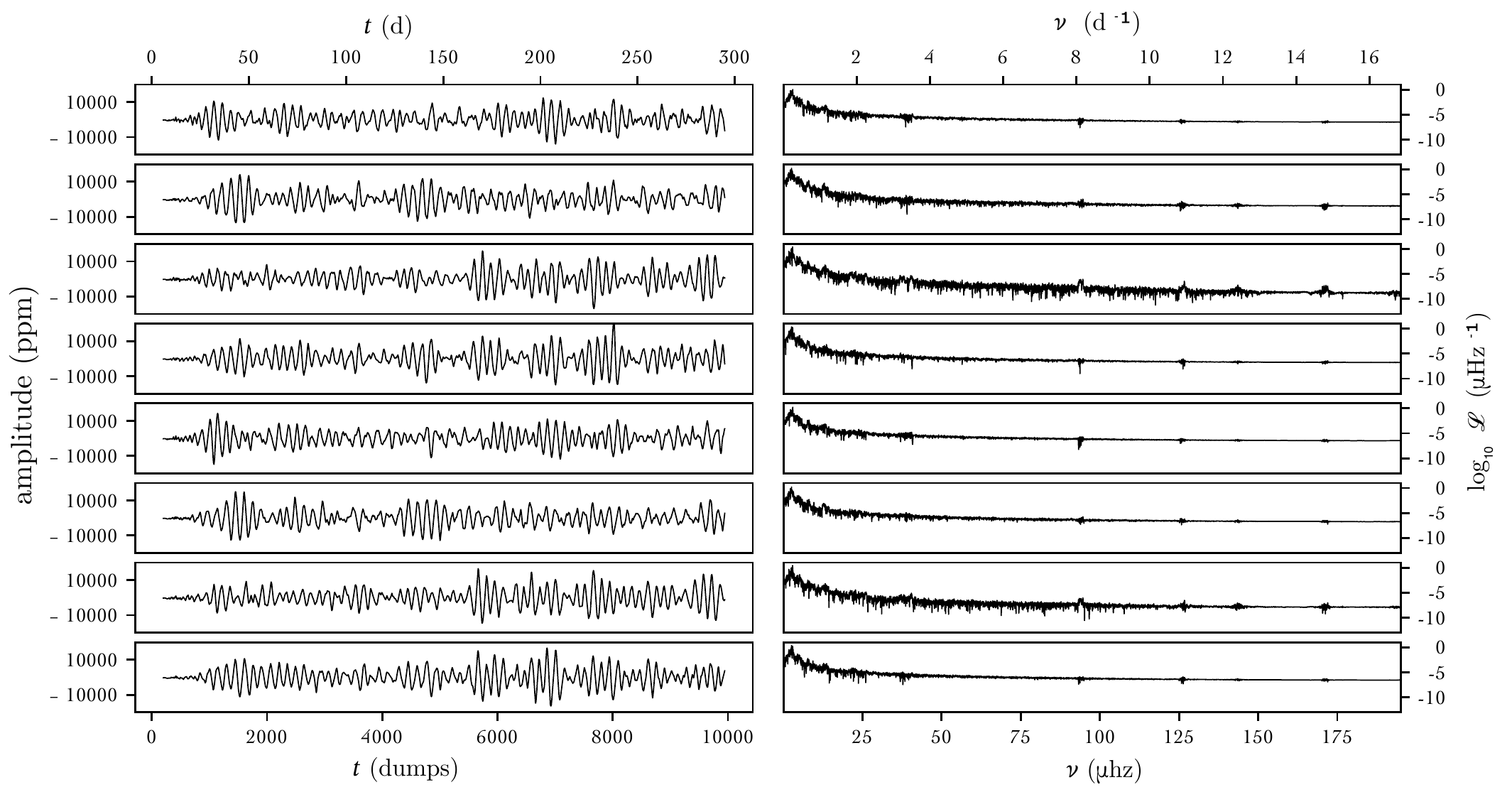}
\caption{From top to bottom, eight simulated light curves (left panels) and their Fourier transforms (right panels) for different lines of sight of the simulation M107. The light curves are hemispheric integrations as described in \Sect{astero-analysis}. For these measurements, a radius corresponding to
  $\unit{19.5}{\Msun}$ was treated as the surface. The light curves have been
  de-trended by dividing out a quadratic fit
  (\Fig{detrend}). The time unit along the bottom horizontal axis is \emph{dumps}, which are written out with a frequency of $\frac{1}{\unit{43}{\minute}}$.
  The \emph{amplitude} axis plots the unity-subtracted luminosity divided by the fitted quadratic trend, in order to represent the change in luminosity amplitude in parts per million. The variable
  $\mathcal{L}$ is the power spectral density of the unity-subtracted, relative luminosity.
  Oscillations are plainly visible in first
  column. The second column, shown on a logarithmic scale, shows the
  distribution of power by frequency. The first \unit{36}{d} of the
  simulation are are part of the initial transition period and omitted, see discussion in \Sect{stoch-excit} regarding the inital phase of the simulation.} \lFig{lums}
\end{figure*}

 We present the results of our simulations and our analysis in the
 following order. First, we describe the general morphology of the
 simulations, such as the convection, the interaction with the
 convective boundary, and characteristics of oscillations directly
 visible in snapshots of the simulations. Next, we present
 oscillation properties of the simulation run M107 ($768^3$ grid) including light
 curves, luminosity power spectra, and the evolution of modes with time.
 Then, we decompose the oscillations in the simulation M114  ($1152^3$ grid)
 to reveal the breakdown by frequency and angular degree, the radial
 extent of each mode by mass coordinate, and the power spectrum
 responsible for exciting the waves. Finally, we compare the
 dispersion relations of our simulations to the predictions from GYRE
 to identify the nature of the oscillations in our simulations and to
 determine their specific mode numbers.
 We discuss the coherence time of g modes in our simulations based on the spectral line broadening and finally discuss the processes contributing to the formation of a stochastic low-frequency variability in the amplitude spectra of luminosity time series.

 \subsection{General morphology}\lSect{morphology}
 The general flow morphology of the core-convection and envelope-wave
 fluid motions in our simulations are discussed in detail in Section 3
 of \papI. Here we summarize the most important points.
 \Fig{movie_panels} shows images of the vorticity, radial, and tangential
 velocity components for run M115 with $1728^3$ grid. The
 large-scale convective motions in the core are arranged like a giant
 dipole. At the time depicted in these figures, it is aligned from
 the north-east to the south-west direction. The radial velocity
 component shows inflow broadly in the top hemisphere and outflow in
 the bottom hemisphere representing a flow through the centre of the
 simulated star. In the horizontal velocity panel,
 the open (to the north-east) horseshoe-shaped
 tangential velocity arching along the boundary of the
 south-western hemisphere represents the return flow.  We do not
 observe many smaller plumes impacting the
 convective boundary.
 
Instead, as the outgoing flow (panel b, red is positive radial velocity)
in \Fig{movie_panels} impinges on the
convective boundary, it is redirected into a sweeping tangential
motion (panel b) back towards the antipode where opposing flows create a
mutually adverse pressure field. Roughly $2/3$ to $3/4$ of the way
around the boundary, the flow against the adverse pressure gradient
starts to separate from the boundary and curls inwards to form the
inflow of the dipole. 
As it does so, it produces a characteristic
feature when projected in three dimensions like in
\Fig{movie_panels}. We call these boundary-layer separation
\emph{wedges}. They are most clearly seen in the vorticity and
tangential velocity panels just inside the convective boundary in the
north-west and south-east direction.

The vorticity is enhanced inside the wedges indicating additional
instabilities generated due to the boundary-layer separation process,
and it is here that most of the entrainment takes place due to these
wedge instabilities (for details see \papI). We have already
identified and described the entrainment at the top of a convection
zone through the boundary-layer separation mechanism in
high-resolution simulations of He-shell flash convection in low-mass
stars \citep[][especially Fig.\,10 and
  11]{woodwardHydrodynamicSimulationsEntrainment2015}. Movies\footnote{Movies
and images are available at \url{https://www.ppmstar.org}} reveal
more clearly than still images that these wedges launch waves into the stable layers.
The features in the wedges are small compared to the largest convective
scales like the dipole flow.
Movies do not always show the convective dipole flow and the
boundary-layer separation wedges clearly since the dipole axis
migrates around, and may at times appear pole-on. For the images shown in \Fig{movie_panels},
we selected a time where the dipole was aligned with the image rendering plane.

Focussing now on the convective boundary just above the core, ripples
propagate along the interface, appearing to emanate from the flanks of
the wedges. They travel along the boundary in predominantly the same
direction as the flow from the wedges. At the inflow side of the
dipole convection, the boundary waves are traveling in opposite
directions and can be seen hitting each other, and exciting motions
with finer scales. In \papI\ we discussed how the immediate
boundary layer where the peak of $N^2$ is located contains IGWs that have radial velocity components with power mostly
with radial order $n=-1$ and high spherical harmonic degree $l$. Careful
inspection of the shell regions around the radius
$r_\mathrm{N^2-peak}$ indicated by dashed circles in 
\Fig{movie_panels} reveals these waves especially in the 
tangential velocity component along the north-western outer wedge rim,
as well as on the opposite side on the outer wall of the south-east
wedge.
        
Above the convective boundary are patterns of repeating, oscillatory
motion in all velocity components. In the tangential velocity
component these appear as long arcs of variable length and radial
extent of $\approx \unit{120}{\Mm}$. In the radial velocity component
coherent patterns have substantially shorter angular extent than in the
tangential velocity component and they have typically a radial extent
that is twice the radial extent of the arcs in tangential
components. The vorticity naturally shows smaller-scale features of
the motions in the stable envelope. However, the vorticity does not
show the irregular distribution of vortex-tubes that is characteristic
for turbulence and clearly seen in the core. These regular velocity
patterns represent wave motions that are excited by the core
convection, as that is the only mechanism present in these
simulations. These oscillations are predominantly larger and longer
period compared to the fine structures rippling along the convective
boundary.

We return to examine these waves quantitatively in
\Sect{exc-spec}. For a more complete discussion on convection, wedges,
their interaction with the convective boundary, and mixing, see \papI.

\subsection{Photometric variability}\label{phot-variab}
We now present light curves extracted from simulation M107.

\subsubsection{Simulated light curves}\lSect{sec:sim-lc}
The eight light curves corresponding to different viewing angles of
the star (cf.\ \Sect{astero-analysis}), taken at mass coordinate
\unit{19.5}{\Msun} are shown in \Fig{lums}. A prominent
periodicity of approximately \unit{2.5}{\microhz} dominates the
oscillation. This periodicity appears in each \los.  In addition,
the light curves show complex and irregular patterns of fluctuating amplitudes and
additional frequencies can be identified. The oscillation patterns are stochastic and
can be understood as the superposition of waves with a broad and fluctuating spectrum
of modes. Importantly, the contribution of individual modes is
variable leading to fluctuations of the amplitudes of the resulting
total oscillation. The variable nature of the oscillation modes and
the corresponding stochastic excitation of IGWs is discussed in
detail in \Sect{stoch-excit}. The oscillation fluctuations are only
weakly correlated or even uncorrelated for the different directions in
the same hemisphere. Opposite directions are in general more
correlated with each other than different directions in the same
hemisphere.

\begin{figure}
\centering
\includegraphics[width=\figwidth]{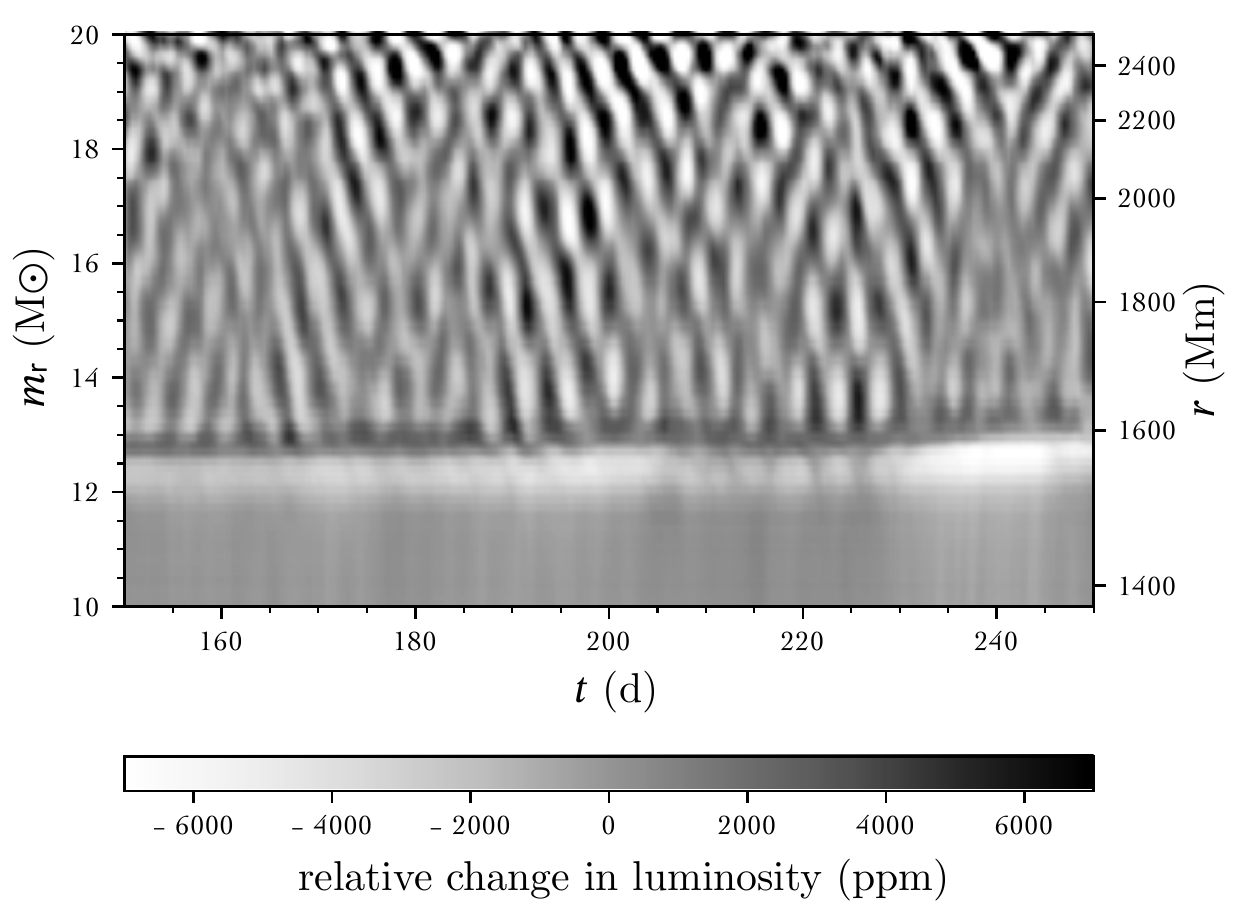}
\caption{The evolution of the simulated photometry as a function of
  radius and time. The convective-boundary region begins at
  approximately $12 \Msun$. Above this, oscillations are visible
  throughout the envelope. By visual inspection of the peaks and troughs
  going outward from the convective boundary, we can identify a prominent
  mode with radial order $n\approx 5$.}
\lFig{lum-evolution}
\end{figure}
The second column of \Fig{lums} shows the power spectrum
corresponding to each time series. The spectra contain a rich
assortment of peaks, with the highest power near 
\unit{2.5}{\microhz}. 
Sharp features near 25, 95, 127, 150, and $170\; \microhz$ visible
from all lines of sight  are p modes (see \Sect{exc-spec} and \Fig{k_omega_all_envelope} for identification).
These spectra also show that there is generally
more power at lower frequencies. Below $\approx \unit{50}{\microhz}$
the spectrum shows more power than above.  Thus, the oscillation
spectra have a low-frequency excess to be discussed in more detail in
\Sect{low-freq-excess}.
The sharp features visible in right panel where it may appear there is a line with a deficit of power are in fact the sidelobes of positive peaks, and are an artifact of plotting the spectrum on a log scale with little no instrumental or photon noise to hide these effects.

Considering only one light curve, we plot the time evolution of the
luminosity as a function of radius in \Fig{lum-evolution}. This
representation emphasizes the stark difference between the regular
wave patterns involving radial scales of $100$ to $\unit{200}{\Mm}$ in
the stable envelope and the turbulent noise in the convection
zone. Since we integrate the light curve over one side of the
simulation, the random turbulent motion averages away. The regular
patterns at mass coordinates $\gtrapprox \unit{13}{\Msun}$ represent standing waves, and again the irregular or
stochastic fluctuation of the wave amplitudes in the envelope is
evident.

\subsubsection{Stochastic spectrum variability}\lSect{stoch-excit}
\begin{figure}
    \centering
    \includegraphics[width=\figwidth]{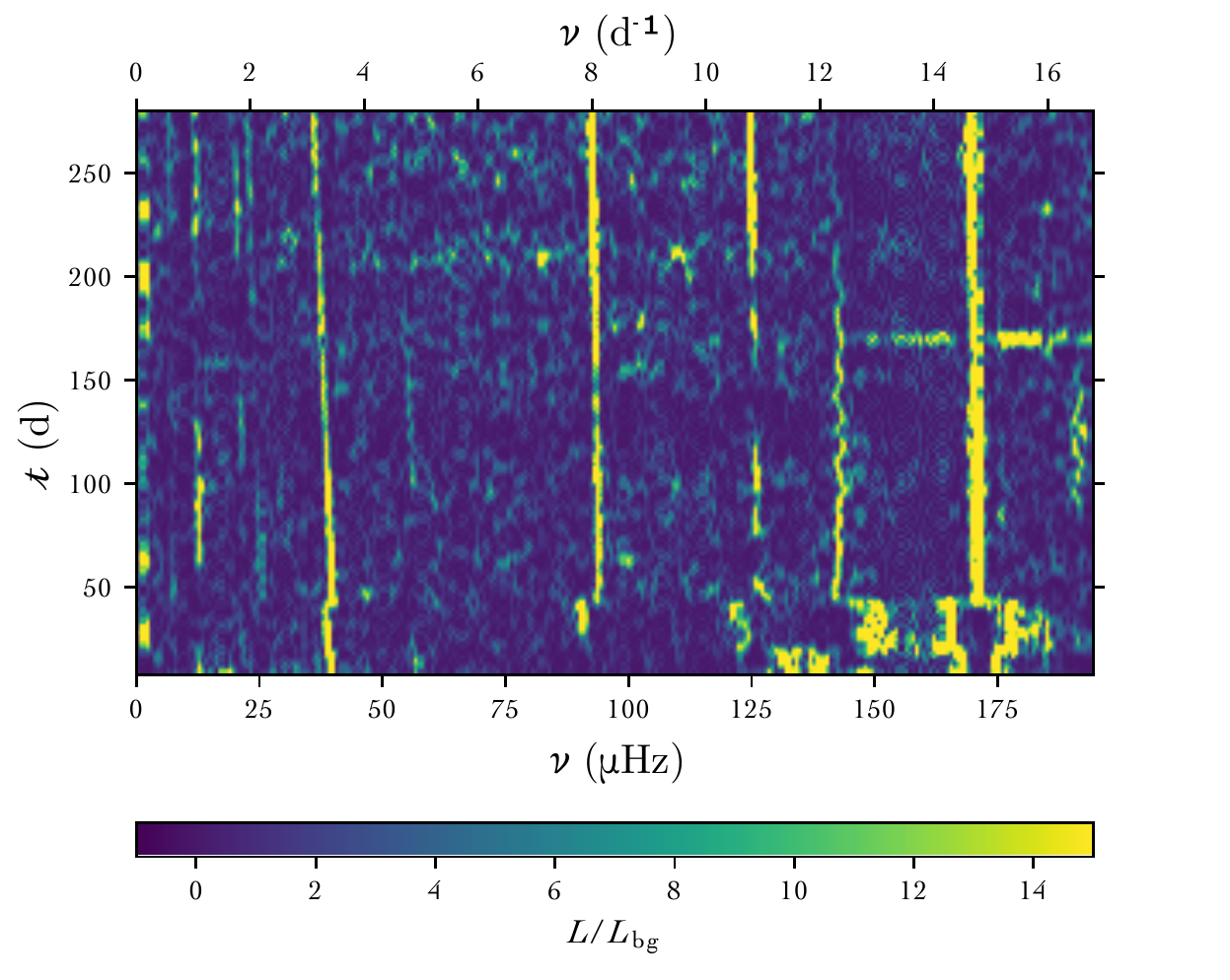}
    \caption{A sliding Fourier transform, or spectrogram, of the
      luminosity time series from one \los. 
      The colour scale shows the luminosity
      power spectral density divided by a smooth background
      created by median filtering the power spectrum. Some modes are
      steady over time, while others are periodically re-excited such
      as the mode at $\approx 2 \microhz$. The initial transient is
      visible in the first $\approx \unit{40}{d}$ of the simulation, which
      are excluded from further analysis. The window size is \unit{15.3}{\days} and we advance
     advance each row in time by  \unit{11.5}{\days}.}  \lFig{sliding_fft}
\end{figure}
In \Fig{lums}, we see an apparently random variation in the power of most modes along each line of sight on the time scale of roughly $\unit{30}{d}$, which we call stochastic variability.
For example, the 6th and 7th line of sight include periods of activity that seem uncorrelated with each other and with those at other times.
  The stochastic variability in integrated light is the result of the superposition of a
spectrum of modes excited sporadically by core convection. This is visible most
clearly in a spectrogram (\Fig{sliding_fft}) which shows the
relative variation of power with respect to a smoothed background
spectrum. Many prominent modes vary in power or frequency over the
length of the simulation.  

During the initial transition period of
approximately one month, the core-envelope boundary ($N^2$-peak) migrates outward from the location defined in the initial setup (see \papI, section 3.1.3 and 4.1 for details). After that
time the $N^2$-peak shape does not change much and only continues to migrate
outward slowly in mass and radius. \Fig{sliding_fft} shows that during this initial transition period the dominant coherent modes are slightly different
than at later times reflecting that the spectrum probes the shape of the $N^2$ profile. We generally
excluded this transition period from our analysis. Note that this initial \emph{transition} period is different from the much shorter initial \emph{transient} time of about one convective time scale(\Sect{conv-freq}) during which the velocity flow field in the convection zone and in the wave region reaches a dynamical steady-state (\papI, section 3.1.3).
         
Power at some frequencies like $\approx \unit{4}{\microhz}$,
periodically fade before being re-excited. Another example of a
somewhat fainter mode can be found at $\approx\unit{21}{\microhz}$
which is at high power from $\approx 110$ to $\unit{150}{\days}$, then
becomes weaker and is excited again at \unit{214}{\days} for a
duration of about \unit{51}{\days}. Our analysis suggests that power
fluctuations shown in \Fig{sliding_fft} on time scale as short as
$\unit{3}{\days}$ are representative of the underlying data (cf.\ \Sect{sec:low-freq-ex-orig} for an estimate of the IGW coherence time based on the spectral line width presented below).

A sufficiently long spectrogram with a short window size, such as \unit{15.3}{\days} in \Fig{sliding_fft} can reveal spectral changes as a function of time, for example due to the migration of the boundary and by
the adjustment of the spherically averaged structure due to heating
without radiative transport. Modes in \Fig{sliding_fft} migrate slightly over the duration of
the simulation. For example, one mode that starts at \unit{50}{\days} and
$39.67\microhz$ shifts to $36.95\microhz$ at the end of the
simulation. We conclude that such dependency of the spectral features on the omission of radiation diffusion and the impact of this assumption on the structure is minor. This is also supported by the GYRE prediction error bars discussed at the end of \Sect{komega}.
        
Stochasticity requires not only a stochastic excitation mechanism but also some form of de-excitation or damping. After the initial transient, the tangential and radial velocity approaches a steady state, as seen in Figs. 7 and 8 of \papI. Since these simulations do not contain the effects of radiative damping, nor significant numerical damping, a possible mechanism for damping or de-excitation in these simulations is that convection is not only a source but also a sink of wave energy. Depending on the phase of the wave and convective motions at the convective boundary where these two types of motions interface wave energy can be returned to the convection zone. In this picture the convective and wave energy content of the simulation assume an equilibrium according to the wave energy excitation and de-excitation efficiencies. 

\subsubsection{Low-frequency excess}
\lSect{low-freq-excess}
\begin{figure}
    \centering
    \includegraphics[width=\figwidth]{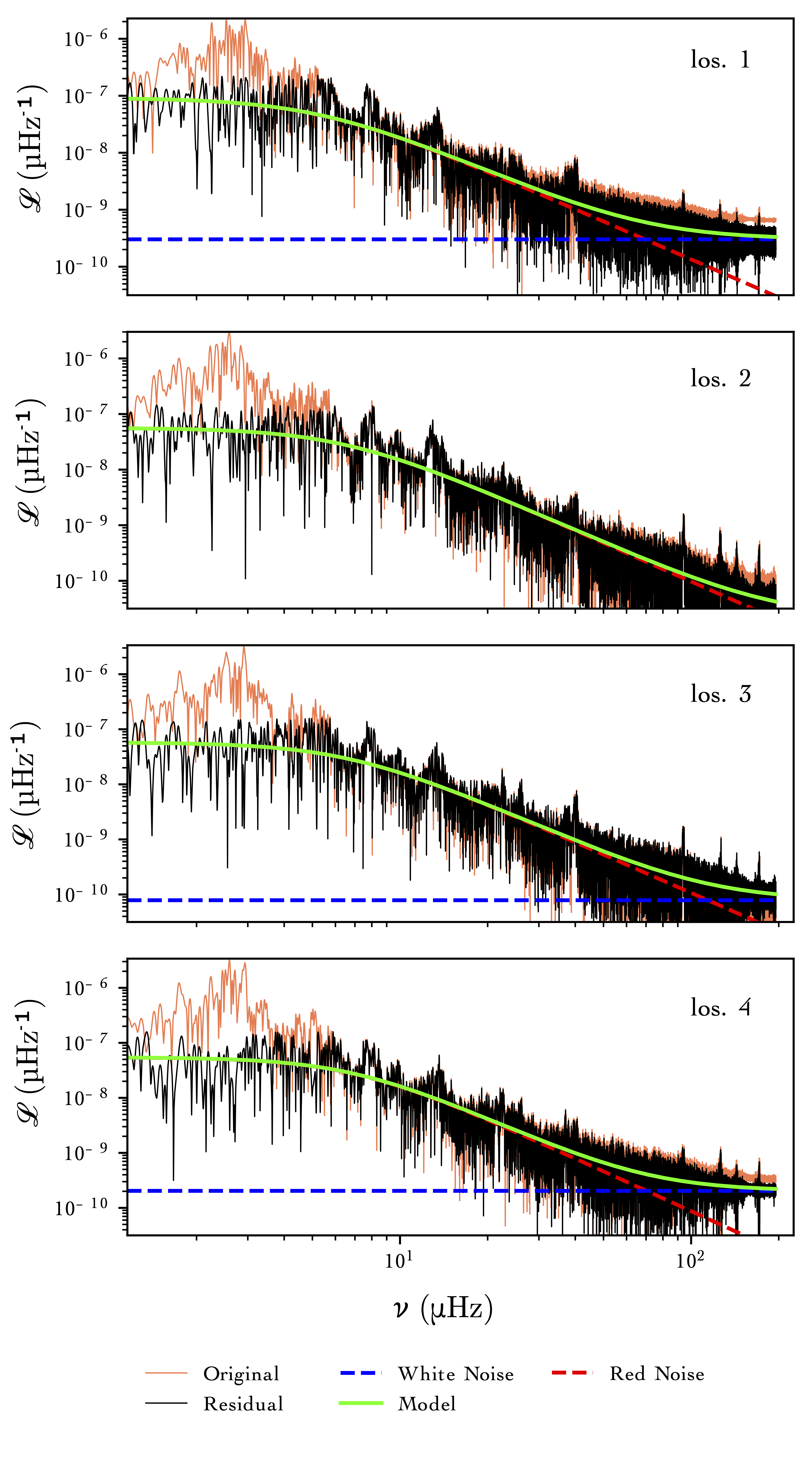}
    \caption{
      Fourier transforms of the photometric variability
      extracted from simulation M107, for four of the eight different
      lines of sight (los.) to the star.
      The variable on the vertical axis, $\mathcal{L}$, is the power spectral density of the unity-subtracted, relative luminosity.
       All time series were
      extracted at mass coordinate $19.5 \Msun$. The orange
      lines are the original spectra, and the black lines are after
      applying iterative pre-whitening. The result is fit to the
      model described in \Sect{astero-analysis}.}
    \lFig{lum_power_spectra_multi}
\end{figure}
\begin{figure*}  
    \includegraphics[width=\textwidth]{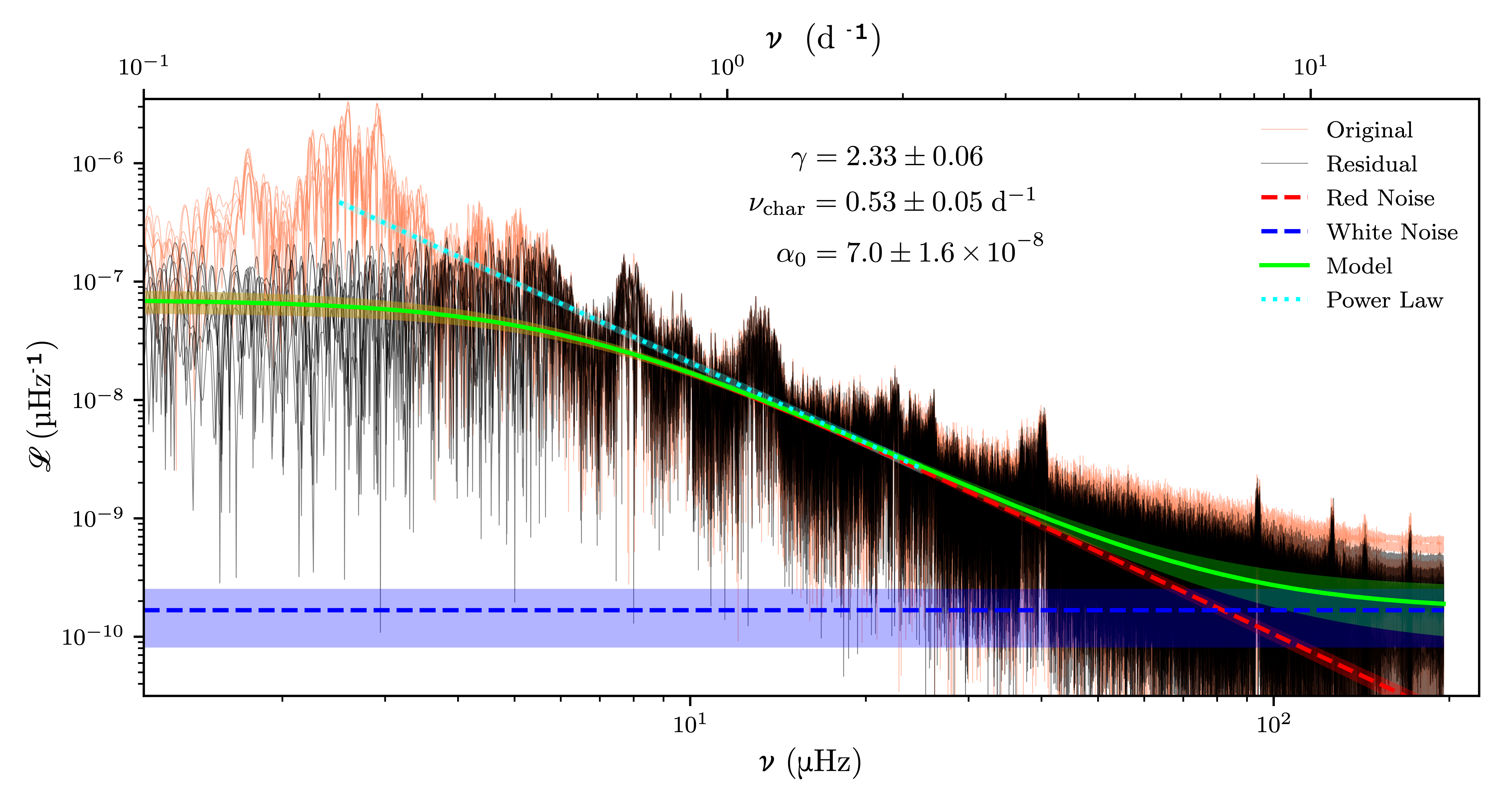}
    \caption{Superimposed luminosity power spectra (M107) of all eight
      lines of sight at a mass coordinate of $19.5 \Msun$. As in \Fig{lum_power_spectra_multi}, $\mathcal{L}$ is the power spectral density of the unity-subtracted, relative luminosity.
      We observe
      an excess of low-frequency power.
      The original spectra
      are shown in orange, and the pre-whitened spectra are shown in
      black. We over plot the mean of the best fitting models of each
      residual spectrum in green, with standard error between
      models marked by the shaded regions. The red and white noise
      components of the model are separately plotted in red and blue
      respectively. 
      We also show a power law fit to the original, not pre-whitened data in cyan.}
    \lFig{lum_power_spectrum}
\end{figure*}
Taking the discrete Fourier transforms of these light curves leads to
the eight panels in the right column of \Fig{lums}. These are the
power spectra of relative luminosity variations, where the luminosity results from hemispheric integration and applying Lambert's cosine factor. The power
peaks at a period corresponding to a frequency of $\unit{2.5}{\microhz}$ or 
$\unit{0.22}{\days^{-1}}$, the
same period we identified in the time series and which is also clearly
visible in \Fig{sliding_fft}. This frequency corresponds to the
characteristic convective frequency (\Sect{conv-freq}).
Motivated by the observations of a low-frequency power excess by \citet{Bowman:2019ib,Bowman:2019ka,Bowman2020b} we apply the iterative pre-whitening procedure described in
\Sect{astero-analysis} that separates and iteratively removes statistically significant (i.e. $\mathrm{S/N} \ge 4$) periodicities
to reveal the background. In doing so we do not suggest that the spectra obtained in such a way are an accurate representation of the observed spectra. An important difference is, of course, that our simulated spectra represent the conditions deep inside the envelope. However, by applying the same data processing steps as done for observed spectra we hope to gain a better understanding of the different effects that in combination may lead to the observed spectra.

The resulting spectra are presented
in \Fig{lum_power_spectra_multi} for the first four lines-of-sight
pointing which all point into one hemisphere. These log-log plots show both the
original and pre-whitened spectra for each \los,  taken at
\unit{19.5}{\Msun}. The residual spectra obtained in this way show an excess of power at low frequencies with a flat
portion at the frequencies below $\approx \unit{10}{\microhz}$ followed by a power-law slope that appears qualitatively similar to that observed in massive main-sequence stars with the K2 and TESS
missions by \citet{Bowman:2019ib,Bowman:2019ka,Bowman2020b} but shifted slightly towards lower frequencies.
Of course, our simulation does not include the surface or the effects of radiative damping, so we cannot say how the spectra would change towards the surface.
We discuss the
properties of the IGWs that contribute to the formation of this
low-frequency excess once we have introduced the \komega s and IGW
mode identification based on \code{GYRE} calculations in \Sect{modes-gyre}.
    
The qualitative similarity of the luminosity spectrum in the stable envelope of our 3D hydrodynamic simulation and
the low-frequency excess observed at the surface suggest that we determine the fitting parameters of the Lorentzian function used to characterize the
observed data.
This is purely to ease the comparison with observations that have had the same pre-whitening and model fit procedure applied.
We fit the Lorentzian model consisting of the sum of
red noise and white noise components to each of the eight
\los~spectra. The level of the flat white noise components
are expected to differ due to the absence of surface effects and
instrumental noise.  The results of the fit to each \los\ are
over-plotted in \Fig{lum_power_spectra_multi}. \Fig{lum_power_spectrum}
shows the power spectrum that results from superposition of all eight
lines-of-sight along with the best-fit model with the parameters
$\gamma = 2.33 \pm 0.06$, characteristic frequency $\nu_{\mathrm{char}} = \unit{0.53 \pm
  0.05}{\days^{-1}}$ (\unit{6.13 \pm 0.58}{\microhz}), and $\alpha_0 = 7.0 \pm 1.6 \times 10^{-8}$, which are the mean
and standard error of the fit parameters across the eight lines-of-sight to the simulation. 
This uncertainty is only the variation expected due to examining the simulated star from different directions; it does not consider the variation expected between different runs of the simulation.

In observations, the range reported by \citet[][Table A.2.]{Bowman2020b} for O dwarfs $2$ to \unit{9}{d^{-1}} with an average of \unit{4}{d^{-1}} for $\nu_\mathrm{char}$ and $1.4$ to $2.5$ for $\gamma$. The characteristic frequency measured from our simulation
is a factor $7.5$ smaller than the averaged in the main-sequence O stars. 
We do not compare $\alpha_0$, the amplitude of the variations, to observations as there are many factors that could impact the comparison to observations (most notably the radius at which it was measured below the surface and the enhanced heating rate).
Finally, applying the iterative pre-whitening to our simulated spectrum as done for observations removes in our case mostly the power in a cluster of frequencies at and around the convective frequency. While the resulting residual spectrum can be reasonably well represented by a Lorentzian model the original simulation data would be better represented by a straight power-law fit \citep{Horst:2020ds} up to the peak near $\unit{2.5}{\microhz}$. This is illustrated in \Fig{lum_power_spectrum} where a power-law fit in log-space to the non pre-whitened data is able to include the power removed by pre-whitening. 

\subsubsection{Convective frequency}\lSect{conv-freq}
The noted cluster of peaks in the region between $2.14$ and \unit{3.00}{\microhz}
(\Fig{lum_power_spectrum}) and first introduced in
\Sect{sec:sim-lc} near \unit{2.5}{\microhz} shown in \Fig{lums},
correspond to periods of $129.76$ and \unit{92.56}{\hour}. The question
naturally arises how these dominant low frequencies relate to a
characteristic convective frequency. This is not a precisely defined
quantity. It involves identifying typical velocities and length scales
to determine a typical convective time scale that would correspond to
a characteristic convective frequency. However, there are many length
scales and velocity scales in the convection region, and their
relative importance is a function of radius (\Sect{s.exspect} as well as 
Figure 6 and Section 3.1.2 in \papI).  Despite following the dipole pattern on the
largest scales the flow is highly turbulent as can be seen from the
small-scale chaotic and random distribution of the vorticity
(\Fig{movie_panels}).
\begin{figure}
  \centering
  \includegraphics[width=\figwidth]{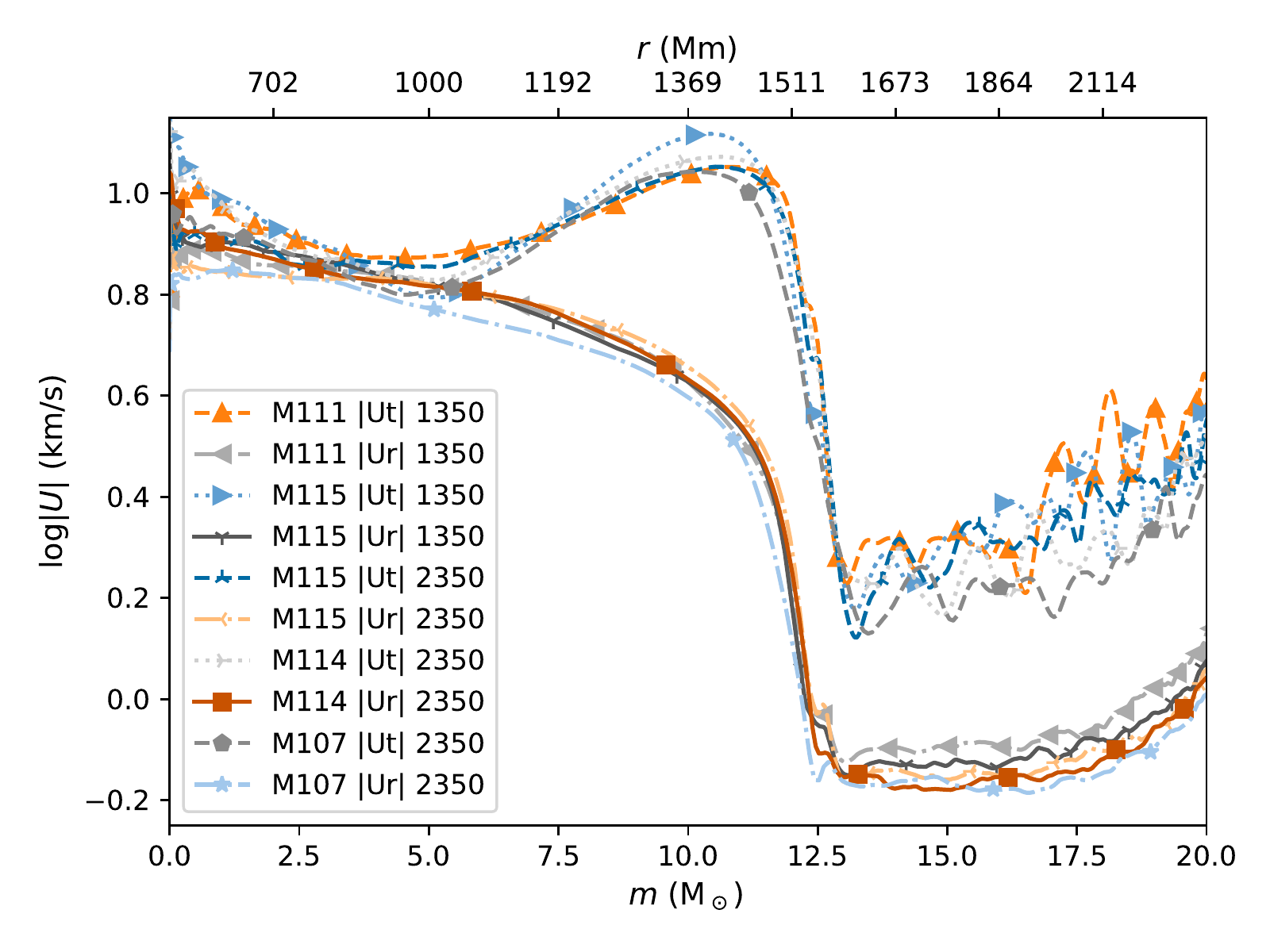}
  \caption{Spherically averaged radial and tangential velocity
    profiles for different resolutions (2688, 1728, 1152, 768, cf.\ \Tab{tab:run_table}). The legend provides simulation name, velocity component and time of profile in \unitstyle{\hour}.}
  \lFig{UtUr-res_U}
\end{figure}

As discussed in \Sect{morphology} the large-scale flow consists of a
central radial column through the centre and a horizontal, arching
return flow along the boundary. This general flow pattern is manifest
in the spherically averaged tangential and radial velocity profiles
(\Fig{UtUr-res_U}). In the centre of the core the radial
and horizontal speeds are the same. Toward the boundary the radial
velocity component decreases starting at a distance of about one
pressure scale height. The horizontal velocity component however
increases toward the boundary and peaks just below the convection
boundary. This represents the comparatively fast boundary-layer
flow. Inspection of the flow images (\Fig{movie_panels} and \papI)
shows that the boundary-layer flow separation typically starts around
$3/4$ of the way from one end of the dipole to the other, (or $3/8$ of
the circumference).
                
From these flow observations we estimate convective frequencies. In
\papI\ we have estimated a convective time scale adopting an average
convective velocity of $U_\mathrm{conv} \approx \unit{6.5}{km/\second}$ and as
a distance the diameter of the convective dipole $2R_\mathrm{conv}$ which is $\approx \unit{3000}{\Mm}$ with $R_\mathrm{conv}\approx\unit{1500}{\Mm}$. This implies a convective time scale of
\unit{128.2}{\hour}, or a frequency of \unit{2.17}{\microhz}. Alternatively one can consider
the circular boundary-layer flow that returns material from where the
radial flow impinges on the boundary to the antipode where the inflow
forms.  Considering that the excitation of waves is in part related to this
boundary-layer flow one may adopt as a length scale the portion of the
flow from where the radial outward-directed flow reaches the
convective boundary and turns around to the location where the
boundary flow separates as $3/4$ of $1/2$ of the circumference, which
is $\frac{3}{4}\pi R_\mathrm{conv}$. The horizontal velocity magnitude
near the convective boundary is  higher ($U_\mathrm{t} \approx
\unit{10}{\km/\second}$) than the average radial velocity (\Fig{UtUr-res_U}). The convective time scale according to this adopted length scale and velocity is
$T_\mathrm{conv}= \unit{98.2}{\hour}$ corresponding to a frequency of
\unit{2.83}{\microhz}.

This exercise demonstrates that reasonably defined characteristic
convective frequencies are in the same range as the prominent features of the low-frequency excess. The direct relation between these
needs to be further investigated through a heating series of 3D
simulations with the same radial stratification but different driving
luminosities. Simulations at 10 times lower driving have $10^{1/3}$
times smaller velocities and therefore a characteristic convective
frequency that is smaller by the same factor. The convective frequency
at nominal ($1000\times$ smaller) luminosity is then $\approx
\unit{0.25}{\microhz}$. Accordingly, if the prominent system of
features evident in the spectrum of these simulations are directly
reflective of the convective velocity they should be found at
correspondingly lower frequency in simulations with lower driving
luminosity. Demonstrating this effect requires very high frequency 
resolution and therefore very long runs. Such an analysis of the heating dependence of asteroseismic simulation
properties \citep[as in][]{LeSaux:22,Blouin:22a} is beyond the scope of this paper. However, spatial spectra for simulations with different heating factors discussed in  \Sect{s.convergence}
 provide partial answers.

\subsection{The nature of the oscillations in the stable envelope}\lSect{exc-spec}
Up to this point we have documented the stochastic properties of the
luminosity oscillations and their correlation with velocity
oscillations. We have established the presence of standing waves with
stochastically variable amplitude. In order to further characterize
the nature of these waves, and in order to determine the origin of the
low-frequency excess of the luminosity spectra at a given radius of
our simulation we construct frequency-wavenumber  \komega s as in \citet{Herwig:2006gk} where it was
demonstrated how these diagrams allow the identification of IGWs and p modes as a function of radius when transitioning
from the convection zone to the stable layer. However, first we 
determine the eigenmodes of the spherically-averaged radial
stratification of the 3D simulations in order to place the \komega s
from 3D simulations into context.

\subsubsection{Predicted modes from \GYRE}\lSect{modes-gyre}
\begin{figure}
    \centering

    \includegraphics[width=\figwidth]{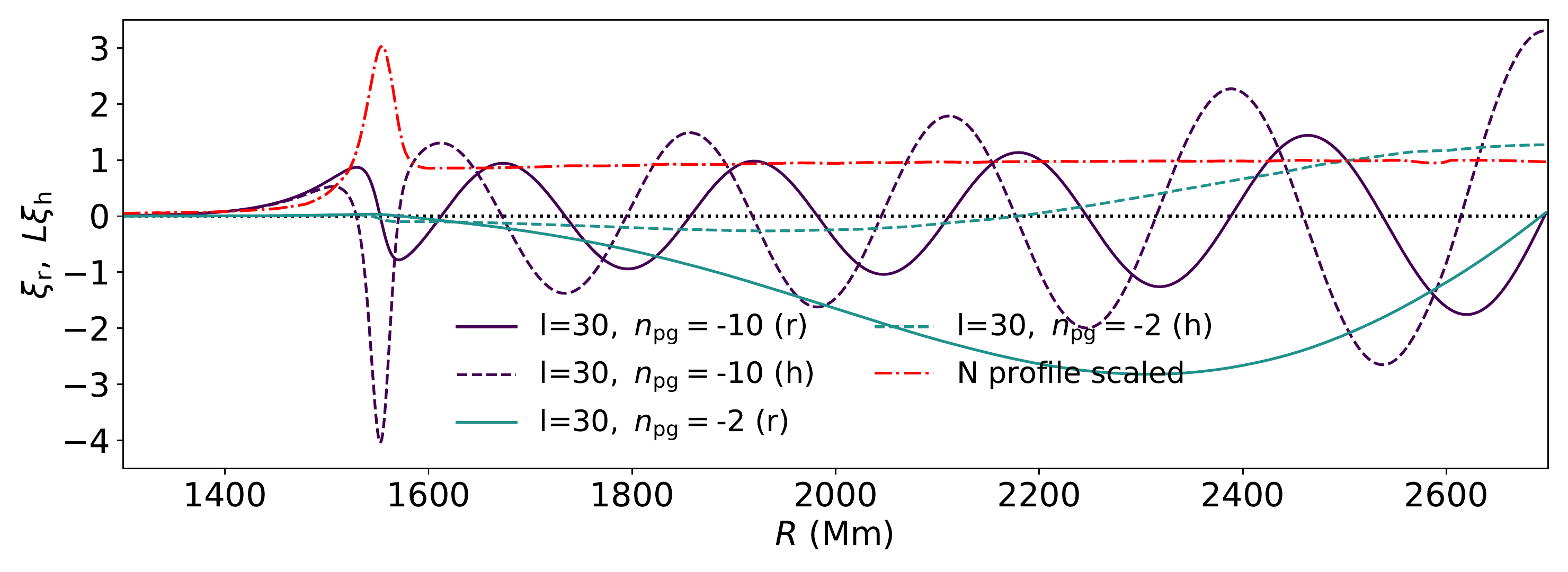}
    \includegraphics[width=\figwidth]{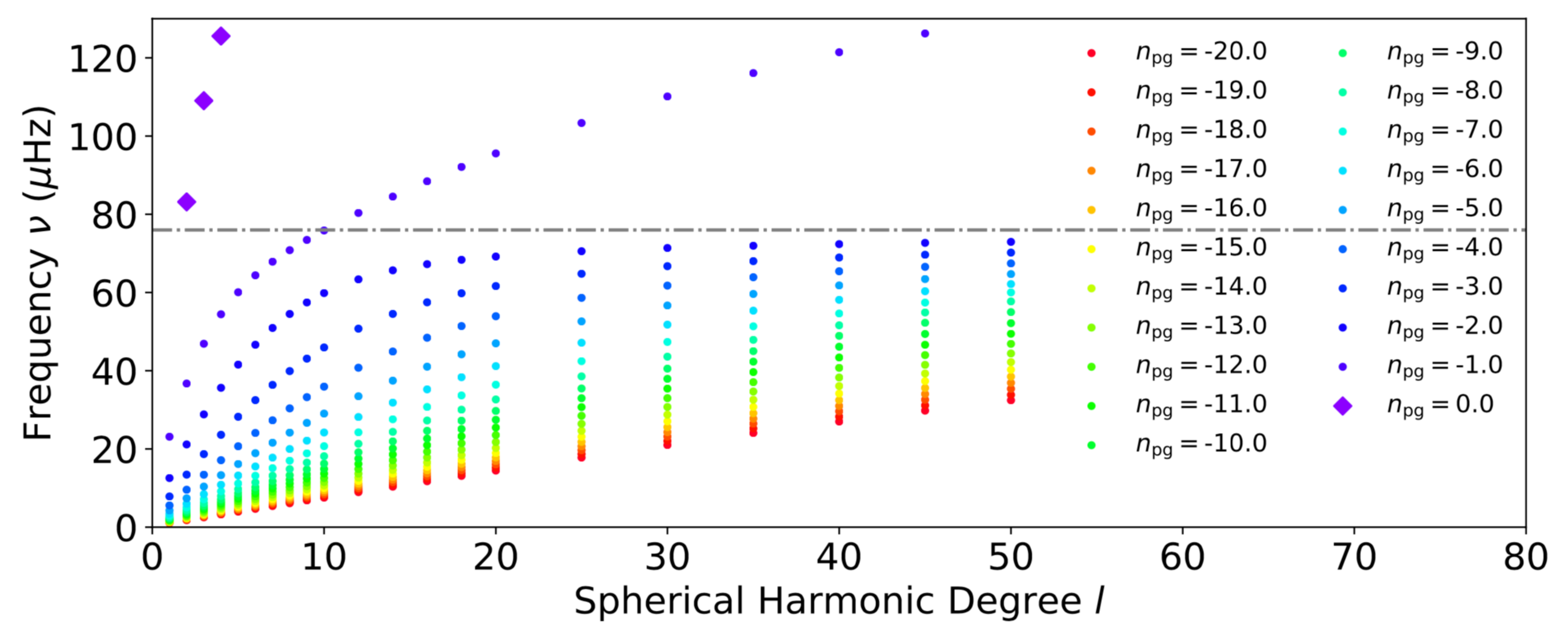}
    \caption{Top panel: Waveforms in terms of the radial $\xi_r$ and
      horizontal $L\xi_\mathrm{h} = \sqrt{l(l+1)}\xi_\mathrm{h}$
      oscillation displacement amplitudes calculated using \GYRE\ on
      spherically averaged radial profiles of simulation M114 at
      \unit{116}{\days} for g modes with $l=30$ and $n=-2,-10$. The Brunt--V\"ais\"al\"a frequency is shown in
      red to indicate the location of the convective boundary region
      in which the $n=-1$ modes resonate (\papI). Bottom panel:
      \komega\ calculated with \GYRE\ for the same radial profile from
      run M114. The grey dash-dotted line shows the \brunt\ frequency representative for the envelope (\Fig{brunt}). $n_\mathrm{pg}<0$ represent g modes. Low
      $l$ f modes for $n_\mathrm{pg}=0$ appear as well. p modes
      with low $l$ have frequencies larger than shown in this plot.}
    \lFig{gyre_modes}
\end{figure}
Usually \GYRE\ is used to determine eigenmodes of a structure
calculated by a stellar evolution code, such as the
\mesa\ code. Instead we use the spherically-averaged radial profile of
a dump in the middle of the range used for the Fourier analysis of the
3D simulation M114 as input for the \GYRE\ code.  This allows us to
identify the eigenmodes we expect in the 3D simulation
(\Sect{gyre-meth}).

\Fig{gyre_modes} shows the
\GYRE\ results \citep[cf.\ ][Fig.\ 2]{Pedersen:2018ew}. In \papI\ we have already discussed the role of the
$n=-1$ g mode that dominates the $u_\mathrm{r}$ power in the narrow
region of the $N^2$ peak where its prominent wave motions are
clearly distinguishable in the velocity image renderings (Fig.\, 18,
\papI). Here we show the waveforms of some higher-radial order modes
that have their largest amplitudes in the stable envelope above the
convective boundary region. By identifying systematically the
frequencies and $l$ for $-20 \leq n \leq -1$ we arrive at the
\komega\ schematic of eigenmodes shown in the bottom panel of
\Fig{gyre_modes}.  As expected g modes have frequencies in the
envelope up to the \brunt\ frequency. For each radial order $n$ we
obtain the arc-shaped dispersion relation that is characteristic of g modes.

\GYRE\ can tell us $\nu$ and $l$ for each eigenmode with radial order
$n$, but not how much power is in each of these eigenmodes. This
information is of course contained in the output of our 3D simulation,
and we will extract it in the next section.

\subsubsection{\komega\ from the 3D simulations}\lSect{komega}
The \komega s extracted from the 3D simulations are based on the last
2000 dumps ($\approx 1300 \mathrm{hrs}$) of the $1152^3$-grid run
M114. We choose to analyse a subset just long enough to achieve
sufficient frequency resolution while limiting blurring the dispersion
relations due to the slow evolution of the boundary in our simulations
(\Fig{brunt} and \papI). 

In \Fig{k-omega-19}, we present
diagrams from six mass coordinates from inside the convection zone and
near convective boundary as well as from several locations in the
stable envelope.

The power by mass coordinate
and frequency is also plotted separately for 21 individual $l$ values
in \Fig{per-l-ur}.
\begin{figure*}
    \includegraphics[width=\textwidth]{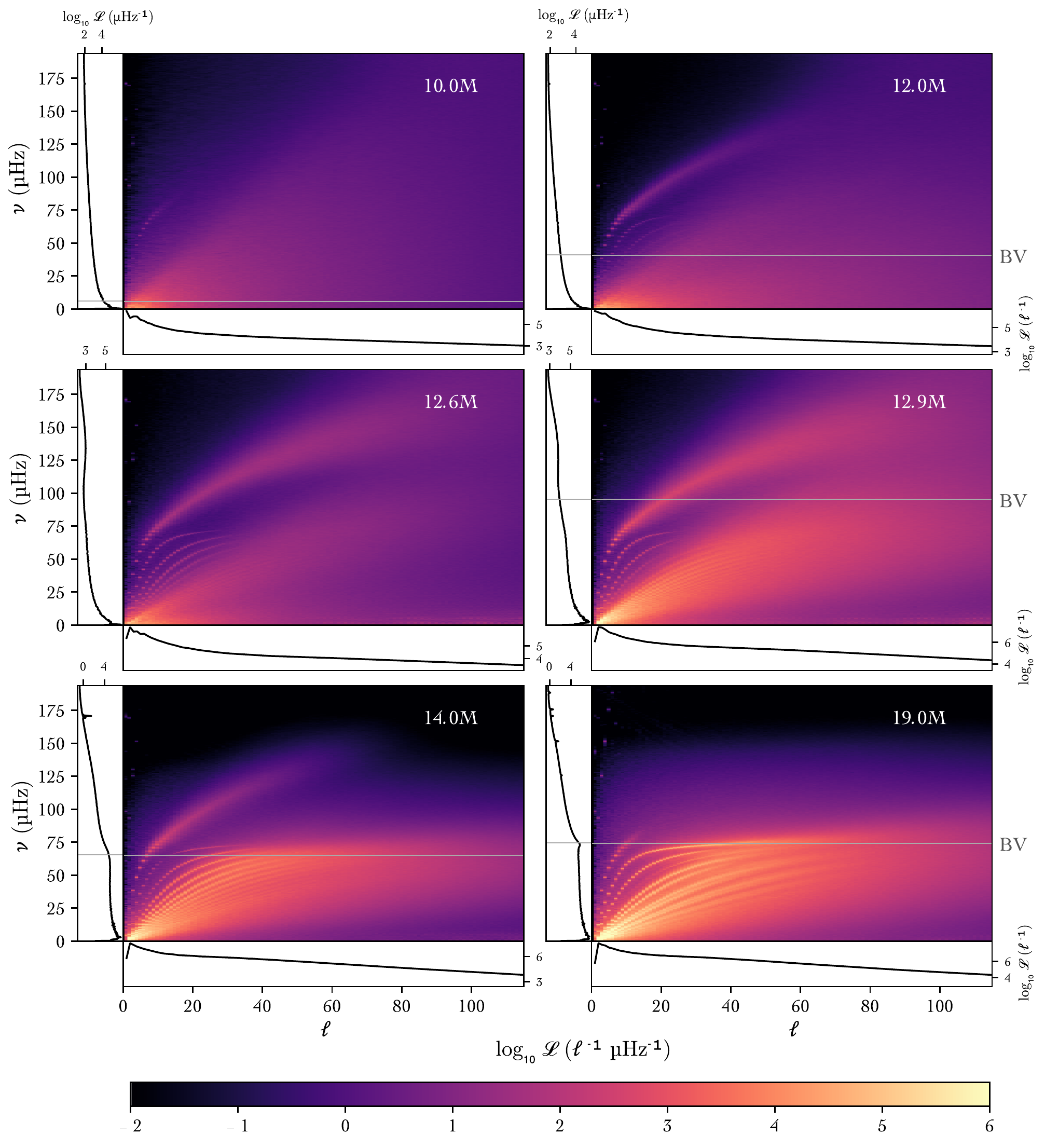}
    \caption{
      Oscillation power spectral density of relative luminosity
      decomposed by both frequency and spherical harmonic for run
      M114.
      The variable 
      $\mathcal{L}$ is the power spectral density of the unity-subtracted, relative luminosity.
      The six panels correspond to different mass
      coordinates. Frequency in $\microhz$ is plotted along the
      vertical axis, spherical harmonic degree $l$ along the
      horizontal, and power is represented logarithmically by the
      colour scale. The plots along the margins show the sum of the
      power along each axis. The horizontal line indicated by
      \emph{BV} gives the Brunt--V\"ais\"al\"a frequency at that mass coordinate.
      Beginning from top left: at $\unit{10.0}{\Msun}$ the power inside the core is smooth and characteristic of
      turbulence, with few distinct features.
      At $\unit{12.0}{\Msun}$ which is
      inside the convection zone just below the convective boundary, the spectrum
      begins to show low order g modes couple to the convection.
      By $\unit{12.6}{\Msun}$, much more of the oscillation power is in a high
      $l$ mode confined to the boundary. By approximately $\unit{14.0}{\Msun}$
      a full spectrum of g modes are excited and the spectrum (vs. frequency and vs. wavenumber) appears flatter out to the local Brunt--V\"ais\"al\"a frequency and $l\approx 60$.
      Further out in the
      envelope such as $\unit{19.0}{\Msun}$, the dominant mode from the
      boundary has died out at high $l$ as it exceeds the local
      Brunt--V\"ais\"al\"a frequency. We cropped the diagrams to only
      show lower frequencies and $l$ values where there is significant
      power and where we expect g modes.
    \lFig{k-omega-19}}
\end{figure*}
\begin{figure*}
    \includegraphics[width=\textwidth]{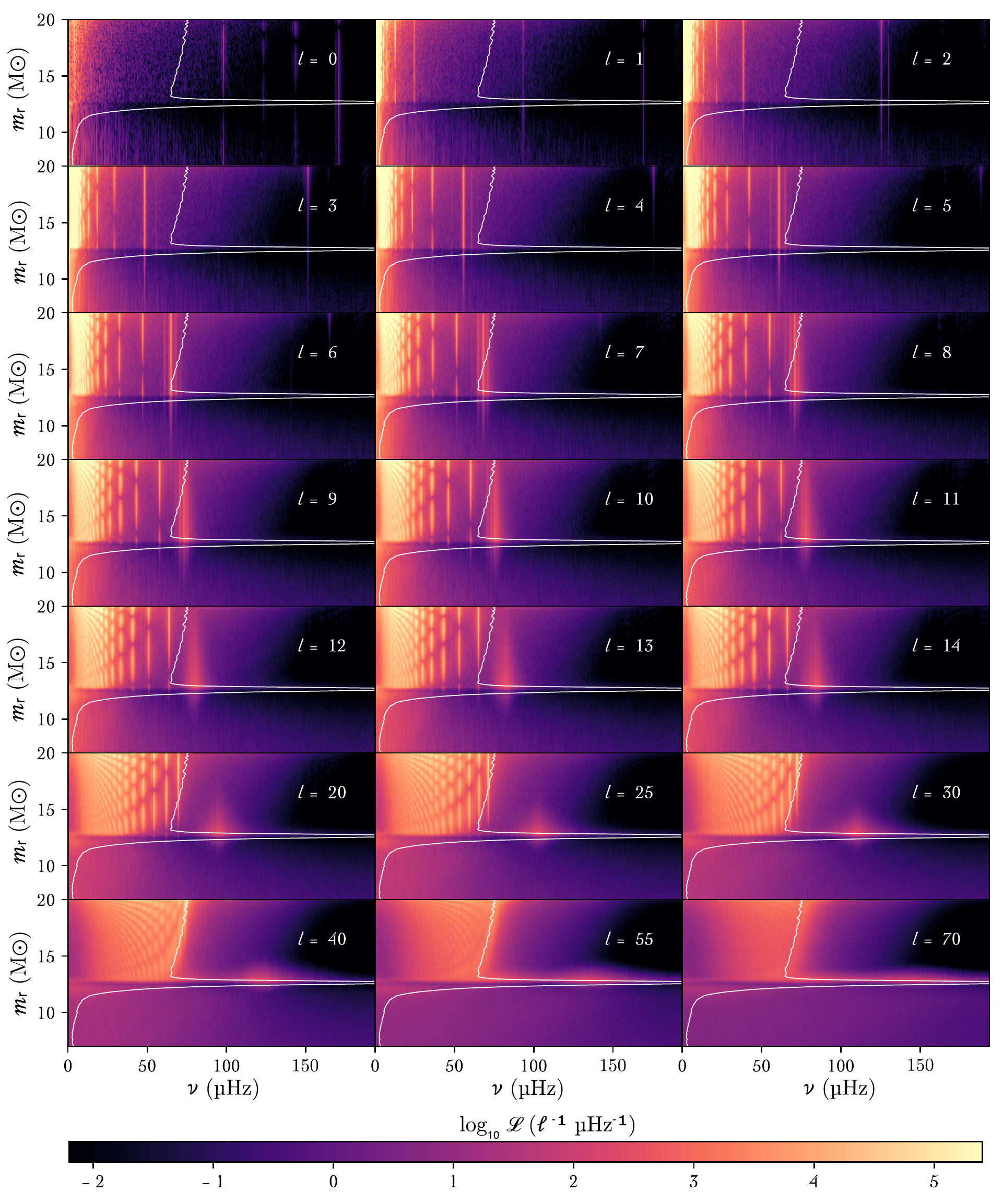}
    \caption{The relative luminosity power spectrum of M114 as a
      function of mass, separated by degree $l$.
      The variable 
      $\mathcal{L}$ is the power spectral density of the unity-subtracted, relative luminosity.
      The turbulence
      visible throughout the core flattens as it approaches the
      convective boundary at approximately $\unit{12.5}{\Msun}$. At this
      point, a forest of g modes are excited in the envelope with
      frequencies approaching the maximum Brunt--V\"ais\"al\"a
      frequency of the envelope. The frequency and number of nodes
      along these vertical lines depend on their radial orders $n$,
      which are related to their angle of propagation through the
      envelope.  The $n=-1$ boundary waves are visible above $f\approx
      60$ peaking inside the narrow boundary region outlined by the
      Brunt--V\"ais\"al\"a profile in white. g modes cannot exist
      with $l=0$, so by process of elimination, the modes visible in
      the top-left panel must be f or p modes, which are not
      considered in this paper.}  \lFig{per-l-ur}
\end{figure*}    
\begin{figure*}
    \includegraphics[width=\textwidth]{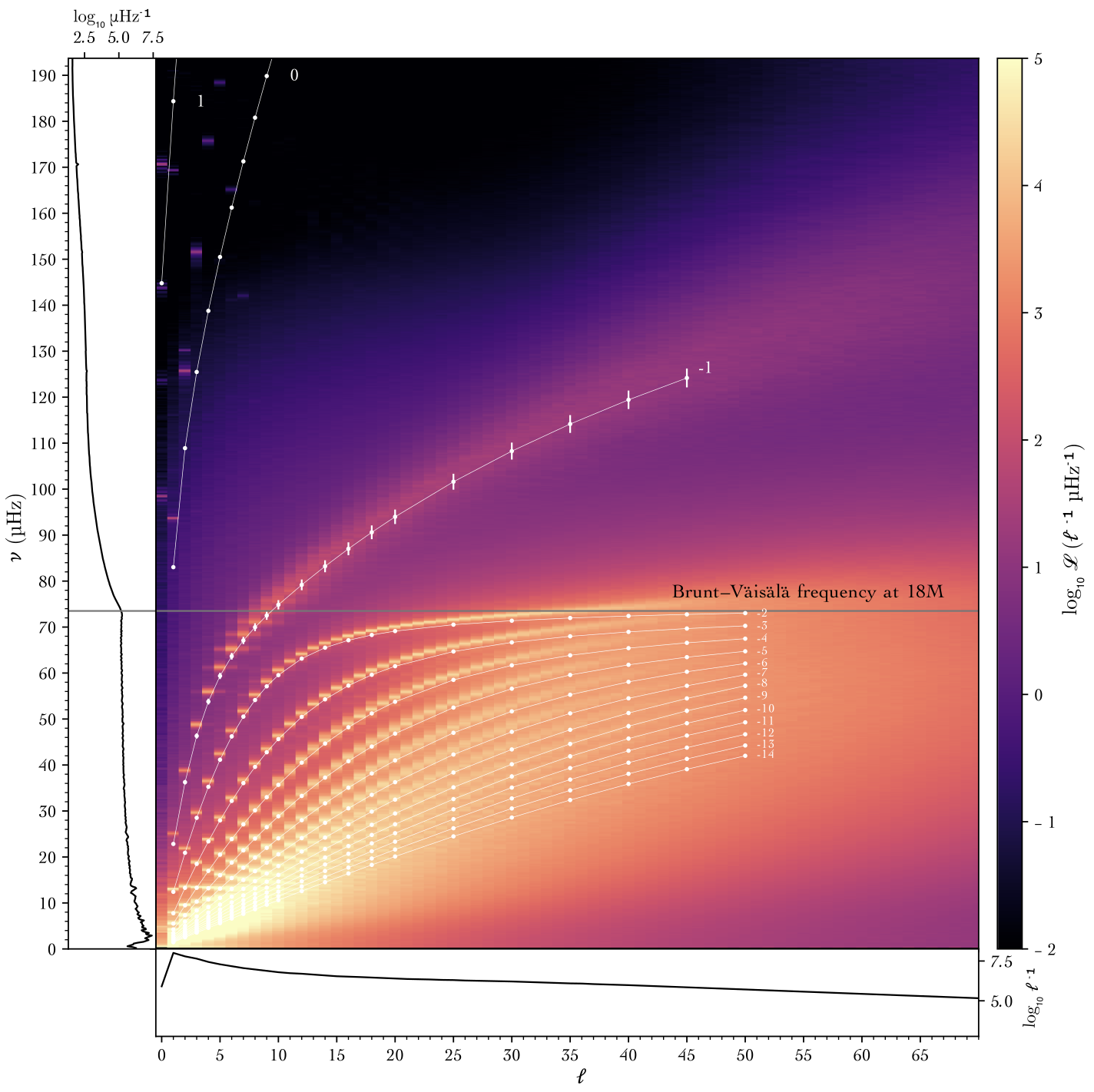}
    \caption{Frequency-wavenumber diagram averaged radially from the
      convective boundary to the edge of the M114 simulation.
      The colorscale displays $\mathcal{L}$, the power spectral density of the unity-subtracted, relative luminosity.
      We over-plot
      the eigenmodes predicted by GYRE as white lines and label them
      by their radial order $\mathrm{n_{pg}}$ where positive indices indicate
      pressure modes, zero indicates the f mode, and negative indices
      indicate g modes. The predictions of GYRE agree with
      the results of our simulations. This
      figure has been cropped -- the numerical resolution of the simulation grid permits modes up to $l\approx1800$. The GYRE calculations were run
      against the start and end dumps of the time span used to create
      the diagram, and then averaged. When visible, the vertical white
      bars indicate half of the difference between those
      calculations.
      We do expect some aliased p modes to visible in top left of the plot.}  
      \lFig{k_omega_all_envelope}
\end{figure*}

In \komega s gravity waves follow dispersion relations that appear as
arcs that asymptotically approach the local \brunt\ frequency (see
\Sect{modes-gyre} for the corresponding \GYRE\ predictions). The
dispersion relations are modified by the radial order $n$, which
relates to the angle of propagation, with oscillations in low $n$
waves being nearly radial and in high $n$ waves with the same $l$
approaching horizontal.

Beginning in the core at $\unit{10.0}{\Msun}$, the convection and turbulence
produce smooth spectra (\Fig{k-omega-19}). The projection onto the
spherical-harmonic degree $l$ horizontal axis shows power to be culminating
at the lowest $l$ values and frequencies without any particular
structure. This is typical for the irregular, intermittent turbulent
motions of convective flow and has been documented already for deep
interior simulations of He-shell flash convection \citep[][Fig.\ 23,
  panel $y=\unit{4.70}{\Mm}$ and $y=\unit{7.45}{\Mm}$]{Herwig:2006gk},
for O-shell convection \citep[although less clearly][Fig.\ 7, middle
  panel]{Meakin:2007dj} as well as core convection
\citep{Rogers:2013fl}. The accumulation of power in
the convective region at low frequencies and low wave numbers reflects
the nature of the turbulent spectrum of convection (e.g.\ Fig.\ 6,
\papI) and is most clearly displayed for radii sufficiently deep
inside and distant from the convective boundary. The power in the
lowest frequency bins of a few $\microhz$ relates to the convective
frequency (\Sect{conv-freq}). However, at $\unit{10.0}{\Msun}$ which is approximately
one pressure scale height below the convective boundary
(\Fig{UtUr-res_U}), a faint signature emerges of the ordered
arc-like pattern of power distribution associated with IGWs
appears. 
This arc belongs to the $n=-1$ modes that have their largest displacement amplitude near the
convective boundary.

Stepping outward, the general pattern is that power redistributes into
different sets of IGW eigenmodes. The mass coordinate $\unit{12.0}{\Msun}$ is
close to the convective boundary on the unstable side. The
distribution of power shifts to a flatter $l$ spectrum. The
arcs of the lowest radial order $n$ IGWs now contain
more power. The \komega\ at this mass coordinate
shows a mix of the unordered convective power distribution and the
regular patterns reminiscent of IGWs. Thus, in the outermost layers of
the convection zone motions are due to a mix of IGWs and turbulent
convection. This gradual transition from convection-dominated to
IGW-dominated flow fields is documented in \papI. At
$\unit{12.6}{\Msun}$ we arrive at the peak of the Brunt--V\"ais\"al\"a
frequency profile. Here, the power in the region at low $l$ and $\nu$
which dominated inside the convection zone is momentarily diminished, demonstrating that irregular, turbulent convective motions generally do not reach
this radial location. It also shows that at and above the $N^2$-peak
radius, motions are dominated by IGWs and not, for example, by plume
overshooting.

In \papI\ we showed that the radial location of the $N^2$ peak is an important separation line between the
convection-dominated flow below and the IGW wave dominated flow
above. In this region of the $N^2$ peak significant power is present in the 
$n=-1$ g modes with a large range of $l \lessapprox 80$ (cf.\ Fig.\, 19,
\papI). Higher order g modes exist but have insignificant power. The
power spectrum has now completely changed, peaking at $l\approx70$ and
$\nu \approx \unit{140}{\microhz}$. The dominant presence of the $n=-1$
mode depends on the exact shape of the $N^2$ profile, something that
3D hydrodynamic simulations aim to ultimately determine. This shape is
a combination of hydrodynamic and secular time scale processes
(\papI). If real stars feature a prominent $N^2$ peak at the
convective boundary the $n=-1$ IGW clearly plays a key role in the
physics of convective boundary mixing.

The boundary region ends at approximately $\unit{12.9}{\Msun}$ where the $N^2$ profile 
reaches a local minimum. At this radius power shifts to IGW modes
that oscillate predominantly in the envelope with $n \le -2$ at lower
frequencies, below the local \brunt\ frequency, and lower $l$. This general trend continues for
larger radii. At $\unit{14.0}{\Msun}$, some distance from the boundary region
and the bottom region of the stable envelope, the power in the high
$l$, $n=-1$ boundary waves is greatly diminished as they now exceed the
local \brunt\ frequency. The spatial power spectrum peaks at low $l$
but has significant power out to $l\approx80$ thanks to a large population of modes with a wide
range of $l$ and $n \le -2$. At this point, the spectrum reaches a turning point at frequencies just below the BV frequency with a sharp drop of power
beyond.

At the largest mass coordinate shown, the power in the $n=-1$ boundary
mode has vanished. Essentially all power is in IGW modes with $n
\leq -2$. A full range of g modes are excited for many $n$ and $l$
values exceeding 100. A key feature of the frequency spectrum summed
up for all $l$ shown on the horizontal axis of each panel, is the large
amount of power near the BV frequency. This power originates
from high-$l$ and relatively low-radial order g modes. It becomes clear that in
order to understand the low-frequency excess shown in
\Fig{lum_power_spectrum} the substantial contributions of
moderate-$l$, and high-radial order $n$ IGW modes near the BV frequency is key.

When interpreting the \komega\ panels for mass coordinates in the stable layer, one needs to keep in mind that the \komega s do not show any power where a mode has a radial node, and thus some modes
along the regular arcs are not seen very strongly.  This is clearly seen for
example for the $n=-4$ g modes for the \unit{19}{\Msun} panel in \Fig{k-omega-19} and in
\Fig{per-l-ur} where the radial power distribution is shown as
function of frequency separately for 21 $l$ values. In that diagram
each vertical line in the stable layer corresponds to one radial order
$n$, and the dark gaps in these vertical lines correspond to
nodes. The frequency of the vertical line for each order $n$ increases
from panel to panel with $l$, tracing out the dispersion relation arcs
from \Fig{k-omega-19}.

It is worth noting that these modes are not sharp and have line widths much greater than we would expect from the Hanning window function applied to control spectral leakage.
Some of this broadening comes from evolution of the Brunt--V\"ais\"al\"a profile throughout the simulation as can been seen in the white error bars of \Fig{k_omega_all_envelope}; 
however, the more significant contributor to the line width of these modes is their finite coherence, since the modes are driven stochastically and have a finite lifetime.

Before discussing the coherence of these modes below, we return briefly to the interpretation of 
\Fig{sliding_fft} now that we have established that the waves are indeed IGW. 
The power at each frequency usually does not correspond to an
individual IGW mode, but rather to a superposition of modes with
different radial order $n$ and spherical harmonic degree $l$ as shown
in \Sect{modes-gyre}. However, in order for the power to vary over
time, the contributing modes must be variable. Thus, the variability of
mode amplitudes as the result of their stochastic excitation by the
underlying core convection is also evident from the sliding spectral representation of \Fig{sliding_fft}.

\subsubsection{Internal gravity waves in the stable envelope}

The general features of the power distribution in the \komega s shown
in \Fig{k-omega-19} show a close resemblance of the eigenmodes
calculated by \GYRE\ shown in \Fig{gyre_modes}. In fact, we can
extend the comparison with the modes as shown in \Fig{per-l-ur} to
include the \GYRE\ information. In the top panel of
\Fig{gyre_modes} the waveforms are shown for $n=-2,-10$ and 
spherical harmonic degree $l=30$. The solid lines gives the
displacement amplitude for the radial velocity component and dashed lines
give the horizontal displacement amplitude.

Comparing the \GYRE\ calculations in \Fig{gyre_modes} to the 3D simulations in Figs. \ref{fig:k-omega-19} and \ref{fig:per-l-ur}, we can directly identify the individual eigenmodes at the correct frequencies and see that their power evolves with radius as expected.

Extending this comparison to many modes, \Fig{k-omega-19} overlays eigenmodes calculated by \GYRE\ from the spherically averaged radial profile to the a \komega\ of the 3D simulation, averaged radially across the envelope.
The radial averaging avoids
\emph{missing modes} that have nodes at some radii. The
\GYRE\ g mode predictions and the results of the 3D simulation
agree very well across the whole range of $l$ and $n$. This shows
that fluid motions in the envelope are indeed predominantly IGWs. The
f mode indicated by $n=0$ and first two p modes are also present but with frequencies
approximately 20 percent higher frequency than calculated by GYRE.

The frequencies predicted by \GYRE\ agree within at most 7 percent relative error for high-$l$
g modes with the frequencies of eigenmodes in the \ppmstar\ simulation.
In \Fig{k_omega_all_envelope} we show the frequency difference of 
\GYRE\ predictions based on the spherically averaged radial profiles
of two dumps, one near the beginning of the sequence analyzed to
create the \komega s, and one near the end. This shows that according to the \GYRE\ prediction the small change in the profile contributes little to the broadening of the lines for all but the $n=-1$ modes. 

\subsubsection{Coherence time of modes}
\lSect{s.coherence}
\begin{figure}
  \centering
  \includegraphics[width=\figwidth]{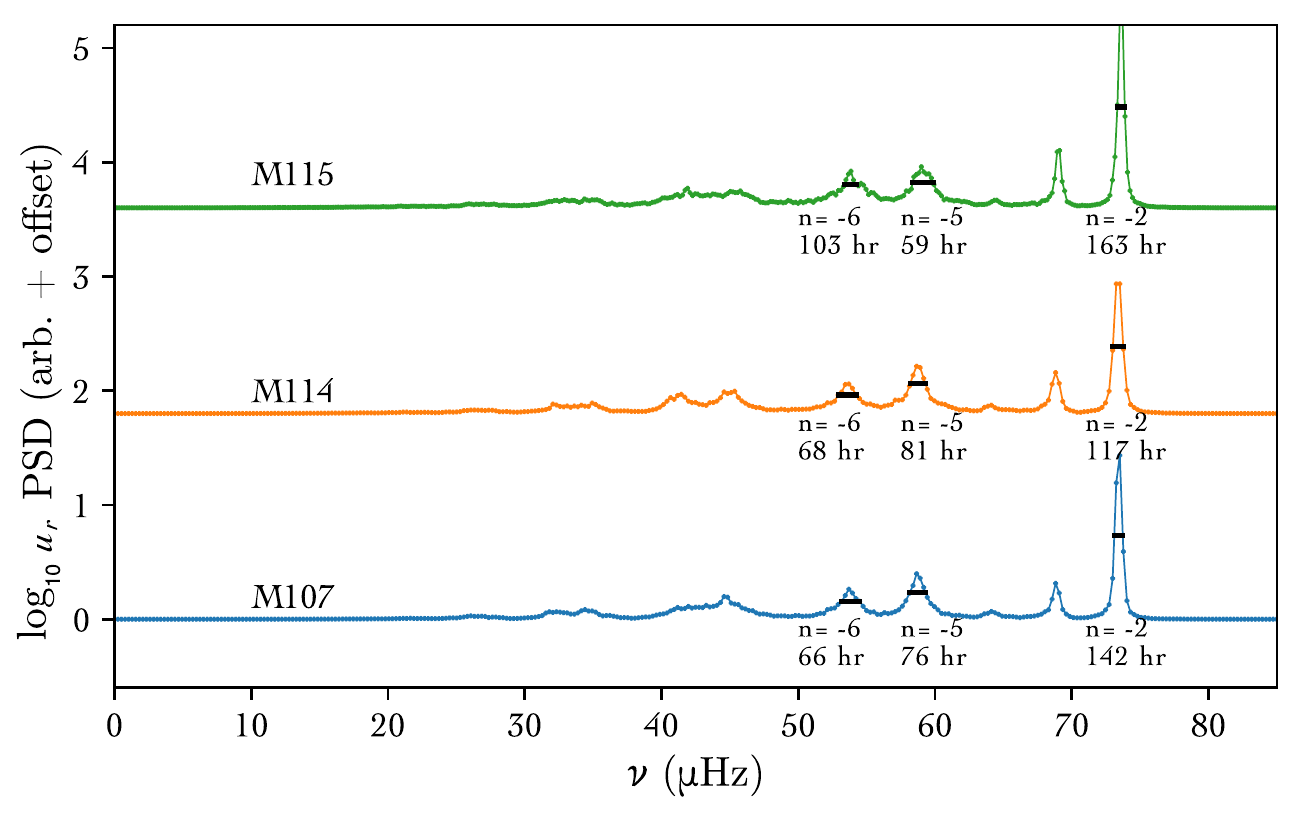}
  \caption{
    Coherence times measured from three modes at $l=30$ and a mass coordinate of $19\Msun$ for three different grid sizes.
    There is no clear trend in coherence time with grid resolution, and considerable variation between $n$ and $l$ mode numbers. 
    The relatively short coherence times are a result of the stochastic excitation by convection
    in the core.
  }
  \lFig{coherence}
\end{figure}
As visible in \Fig{sliding_fft}, we find that the standing g modes in the envelope vary in power considerably over time. To evaluate the coherence of these g modes, we measure the full width at half maximum of a few characteristic peaks in the M114
\komega~in \Fig{coherence}.
The full width at half maximum $2\Delta\nu$ of a mode that is not otherwise broadened by other effects
is directly related to the coherence time by the relation 
$1/(2\Delta\nu)$.
For low-$n$ g modes of order $l=30$, we find coherence times in the range of 80-100 h.
We attribute this limited coherence and short lifetimes to the stochastic nature of the excitation and re-excitation from core convection (cf.\ discussion at the end of \Sect{stoch-excit}).

\subsection{Origin of the low-frequency excess}
\lSect{sec:low-freq-ex-orig}
We are now equipped to describe the origin of the low-frequency excess
(\Fig{lum_power_spectrum}) described in \Sect{low-freq-excess}. By
doing so we reconsider two points that could be raised as possible
challenges for a low-frequency excess having the general shape of the one shown in
\Fig{lum_power_spectrum} to originate from excitation through core
convection
\citep[e.g.][]{lecoanetLowfrequencyVariabilityMassive2019}. One is
that the core convection excitation is limited to low $l$ and the
other related point is that higher-$l$ modes would cancel in the
hemispheric integration. A third important point is the effect of
radiation that we do not include in this paper (see \Sect{discussion}).

\subsubsection{The excitation spectrum}
\lSect{s.exspect}

A key result from our high-resolution 3D core-convection simulation is
a detailed characterization of the dynamic motions near and across the
convective boundary (see \papI\ for additional details). 
As concluded in \Sect{s.coherence}, wave energy is also returned to the 
convective core to arrive at an equilibrium between excitation and de-excitation. For this reason, it might be better to describe this as the \emph{equilibrium spectrum}. 
Regardless of the name, a simple
picture of convection motivated by mixing-length theory would suggest
that the dominant power resides in the lowest $l$ modes (representing
the mixing-length blob rising for the distance of a mixing
length).
While we do see that power peaks at low $l$ modes, significant power
is nonetheless present up to much higher $l$ as well.

We showed in \papI\ (Fig.\ 6) that the
spectrum of the radial velocity flattens significantly near the
convective boundary compared to the regions deep inside the convective
core and compared to the tangential velocity. This is simply a result of the low-$\mathcal{M}$ Mach number and therefore essentially incompressible flow changing from predominantly
radial to predominantly horizontal near the convective boundary
(\Fig{UtUr-res_U}). This flattening of the spatial spectrum near
the convective boundary is a general feature of interior convection
and has also been documented for He-shell flash convection in rapidly
accreting white dwarfs
\citep{stephens3D1DHydronucleosynthesisSimulations2021}.

The motions
contributing to power at large $l$ are physically associated with the
boundary-layer separation wedges (\Sect{morphology}, \papI). Thus, at
the convective boundary a broad spectrum of scales and frequencies
with power at small and large $l$ are present that couple to and inject energy into the
IGW modes of the stable envelope.
Finally, \Fig{per-l-ur} reveals that it
is not strictly necessary to have as much power at a given $\nu$ and
$l$ at the convection boundary as in an excited IGW in the stable
layer.
  
\subsubsection{Attenuation}\lSect{sec:attenuation}
\begin{figure}
\centering
\includegraphics[width=\figwidth]{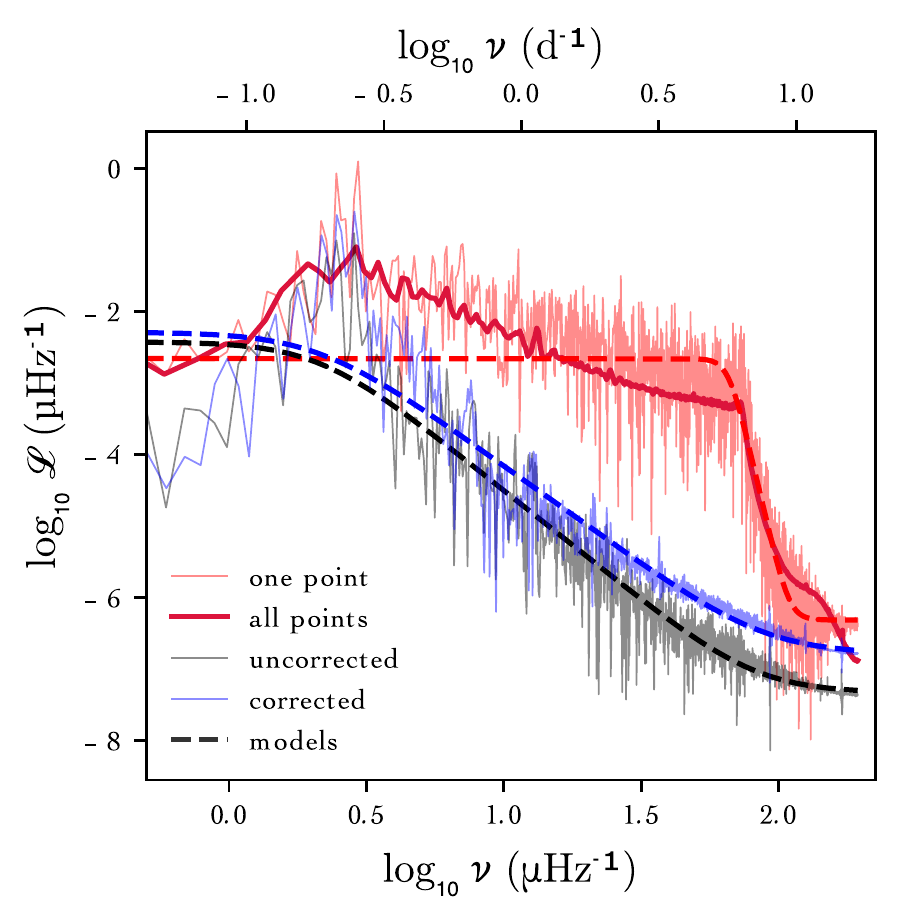}
\caption{ Luminosity spectra extracted at \unit{19.5}{\Msun} from our
  M107 simulation in three different ways. The ``one point'' spectrum
  is measured at just a single point on the sphere. The
  \emph{uncorrected} spectrum is integrated uniformly over one
  hemisphere.  The \emph{corrected} spectrum integrated over hemisphere
  and corrected for Lambert's cosine factor $\cos(\theta)$, converting
  the surface integral into an integral over the projected disk of the
  star.  The integrated luminosity power spectra are attenuated
  significantly where the spectrum is dominated by higher $l$ modes.
  The latter method is used throughout this paper.  The \emph{all points}
  spectrum is similar to the \emph{one point} spectrum, but is instead
  extracted from a frequency-wavenumber diagram of M114 (left margin
  panel of Fig. \ref{fig:k_omega_all_envelope}). This is similar to
  averaging \emph{one point} spectra across a surface of the sphere.  }
\lFig{Lambertian-correction-vs-uncorrected-and-1point}
\end{figure}
\begin{figure}
    \includegraphics[width=\columnwidth]{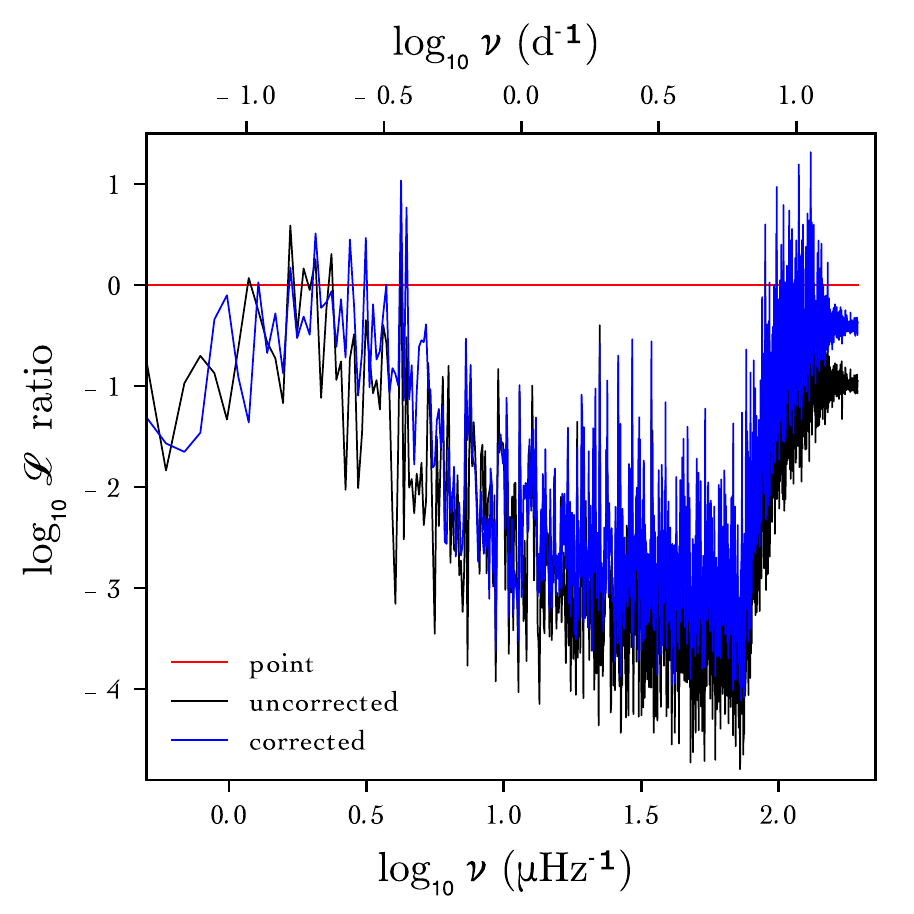}
    \caption{
    The attenuation of the luminosity power spectrum when integrated
    across one face of the simulation, compared to a spectrum
    extracted from a single point (integrated power spectra divided by the single point spectrum).
    The strongest attenuation occurs where the spectrum is dominated by higher $l$ modes.}
    \lFig{integ-methods}
\end{figure}
Through the analysis of the \komega s presented in the previous
section we have established that the fluid motions in the envelope are
IGWs and that the substantial power contribution to frequencies just
below the \brunt\ frequency originates from the superposition of modes with
the approximate range of spherical harmonic degree $15 \lessapprox l
\lessapprox 40$ and with small radial orders in the range $-6  \lessapprox  n
\leq -2 $.
While the highest power lies in individual modes with low $l$ and $\nu$,
the combined power from the
superposition of waves with a wide range of higher $l$ contributes significant power out near 
the \brunt\ frequency. Radial order $-5 \leq n \leq -2$ and $l \approx 30$
modes  are responsible for the power in the upper half of the
frequency range up to the BV frequency.

We now consider how different approaches of turning the 3D wave field into a simulated light curve impact the resulting synthetic spectrum. We consider three options. The first is to take the luminosity only at one point on a 3D sphere of a given radius, somewhat similar to \citet{Edelmann:2019jh} who used eight points. We refer to this in \FigTwo{Lambertian-correction-vs-uncorrected-and-1point}{integ-methods} as \emph{one point}. 
The second is to integrate at each dump over a hemisphere of the simulated star at a given radius and into the direction of a given line-of-sight (\Sect{astero-analysis}) to obtain time series of single luminosity values. Finally, we consider the correction due to Lambert's cosine factor as described in \Sect{astero-analysis}. The latter two options are referred to as \emph{uncorrected} and \emph{corrected} respectively.
The spectra generated using these three options are compared in 
\Fig{Lambertian-correction-vs-uncorrected-and-1point}. A tentative
fit of a Lorentzian model to the \emph{one point} spectrum places $\nu_\mathrm{char} \approx
\unit{3.8}{\days}^{-1}$ close to the BV frequency of this model, and much higher than the value determined for the spectrum according to the \emph{corrected} approach reported in \Sect{low-freq-excess}.
The \emph{one point} spectrum has a pronounced
flat low-frequency plateau with a relatively gentle slope from the peak near the convective frequency toward 
the envelope BV frequency.

That said, a photometric observation receives light that is the 2D
integration over the projected face of a hemisphere. This integration
cancels luminosity peaks and troughs of oscillations. The
cancellation is expected to be largest for high-$l$ modes, and for
coherent oscillations only the partial peaks and troughs along the rim
that have no exact counterpart are expected to contribute net flux to
the oscillation. This rim contribution is further diminished by
applying the $\cos (\theta)$ Lambertian projection factor
(\Sect{astero-analysis}). These are all good reasons to expect that
high-$l$ modes do not contribute significantly to spectra over an
entire hemisphere. However, our simulations suggest otherwise.

The \emph{uncorrected} spectrum in \Fig{Lambertian-correction-vs-uncorrected-and-1point} also shows the
spectrum of the same simulation, but integrated over a full
hemisphere. The cancellation effect of the integration is clearly
present, and it attenuates power from the higher frequency range below
the BV frequency that is associated with high-$l$ modes (as explained
in the previous section). Performing a Lorentzian model fit results in
a substantially lower parameter $\nu_\mathrm{char}$ (which identifies
the edge of the low-frequency plateau) and a smaller parameter
$\gamma$ (that describes the slope from the red noise to the level of
the white noise). However, the power is not attenuated completely 
and, at least in the simulation, we still detect power out to the BV frequency,
Thus, the hemispheric
integration indeed cancels high-$l$, but not to the level that
they become entirely insignificant.

Applying the $\cos (\theta)$ correction further
reduces $\nu_\mathrm{char}$, though the additional effect is relatively small. This can
also be seen in \Fig{integ-methods} where we show again the
spectra for all three cases as well as their ratios. Power at high $l$
is largely present at intermediate frequencies approaching the BV
frequency of the envelope (see \Fig{k-omega-19}). Accordingly, the
attenuation between considering just one point on the equator and
integrating over one face of the simulation is strongest between
frequencies of 1 and 10 $\unitstyle{\days}^{-1}$. Applying the
$\cos{\theta}$ correction does not effect the cancellation
significantly. 

This suggests that the oscillation contributions
escaping cancellation do not primarily come from the rim, but that
instead cancellation effects are smaller than one may expect because
the oscillations are stochastic
(\Fig{lum-evolution},\Sect{stoch-excit}, \ref{sec:sec:low-freq-ex-orig}). Especially stochasticity
in phase would reduce cancellation effects compared to the case of
coherent oscillations. In fact, we have seen in \Fig{lums} that
the luminosity oscillations have fluctuations in
different directions of the same hemisphere with different phases.

This leads to a key result. A power spectrum such as the one shown
in \Fig{lum_power_spectrum} is the result of attenuation through
hemispheric integration of high-$l$ modes of an underlying spectrum
relatively flat for $\nu < \nu_\mathrm{BV}$. The attenuation does not
entirely suppress the modes with frequencies closer to the BV
frequency. Instead, the slope of the decrease of power from the red
noise level at $\nu_\mathrm{char}$ to the intercept of white and red
noise changes by the cancellation due to hemispheric integration. This
slope is expressed in terms of the parameter $\gamma$ of the
Lorentzian model fit.
It in turn depends on the power
distribution in terms of $l$ and $\nu$ and on the stochastic nature of
the excitation, as we believe this reduces the cancellation compared to the case
of coherent oscillations.

Finally, we note that the low-frequency power excess falls to the level of the white noise at the BV frequency. In other words, the location where the red noise intercepts the white noise indicates the peak BV frequency in the envelope, assuming that no other observational white noise sources dominate. This frequency $\nu_0$ changes less than
$\nu_\mathrm{char}$ when integrating over the hemisphere with and
without Lambertian correction factor.

\subsection{Convergence and dependence on heating factor}
\lSect{s.convergence}

For results of any 3D hydrodynamic simulation an important question is always to what extent the finite grid resolution affects the results in question. 
This is especially true in light of conflicting results between simulations, and how those results depend on luminosity and resolution \citep{lecoanet:2023}.

\begin{figure} 
  \includegraphics[width=\columnwidth]{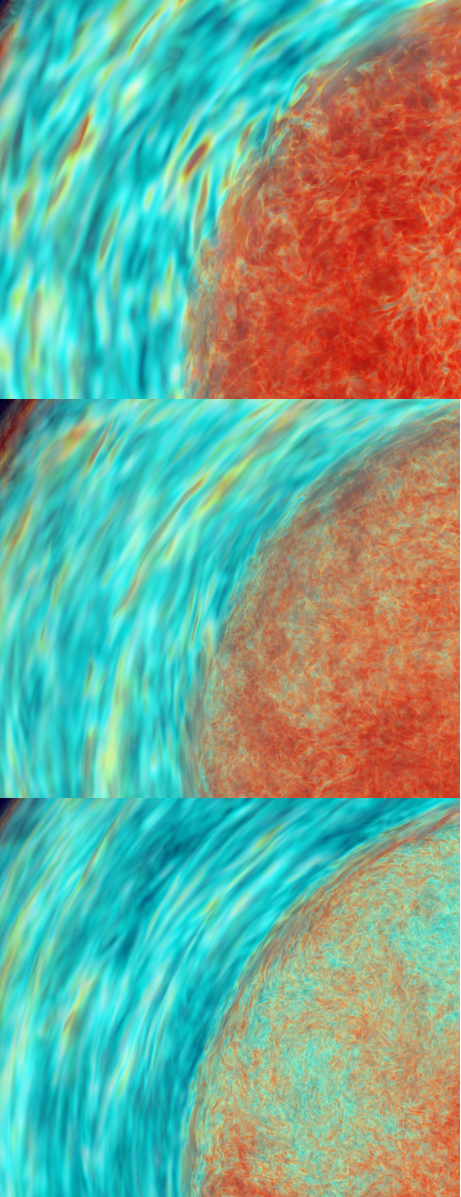}
  \caption{Zoom-in vorticity images (center plane) for dump 2150 (\unit{1540}{\hour}) for simulations M107 ($768^3$), M114 ($1152^3$) and M115  ($1728^3$), from top to bottom.}
  \lFig{Image_compare_resolution_tiled}
\end{figure}
\Fig{Image_compare_resolution_tiled} shows vorticity images for a zoomed-in region of the simulation for three different grid resolutions taken from the same time step. The overall morphology of the flow is similar for all three resolutions and as expected higher resolution simulations show more small-scale structure. 

This is reflected in the spatial power spectra of the velocity magnitude shown in \Fig{convergence-spec} for simulations with grids from $768^3$ to $2688^3$. Note that spatial spectra velocity magnitude are essentially identical to luminosity spectra which were for technical reasons not available for all simulations shown in  \Fig{convergence-spec}.
As expected, simulations with higher resolution contain power at increasingly higher spherical harmonic degree $l$.   Considering the dispersion relation of IGWs (\Fig{k_omega_all_envelope}) high-$l$ power is associated with higher frequencies where most of the attenuation takes place (\Sect{sec:attenuation}). This de-emphasizes the already small impact of grid resolution on the frequency spectra since geometric attenuation is most effective at high $l$. At lower $l$ the agreement between simulations with grids of $1152^3$ and higher is very good. Importantly, the shape of the spectrum depends little on grid resolution.  

The temporal spectra for the same set of grid resolutions is shown in the top panel of \Fig{temporal_power_spectrum}. These have been obtained based on the time series shown in   \Fig{Resolution_lums_time_series}. As can be expected higher resolution runs are shorter. This impacts the frequency resolution. However all simulations are long enough to capture the peak of the power spectral density at $\approx \unit{2.7}{\microhz}$, essentially the same as the convective frequency (\Sect{conv-freq}). We exclude as before the initial transition period (\Sect{sec:sim-lc}) and compared to spectra shown in \Sect{low-freq-excess} we are using for this comparison the same (smaller) radius as for the spatial spectra (\Fig{convergence-spec}). The good agreement between spatial and temporal spectra for the four resolution simulations spanning overall a grid size factor of $3.5$ as well as the absence of an obvious trend of the mode life times reported in   \Fig{coherence}
suggests that numerical effects commonly associated with insufficient numerical resolution are not impacting our results. 
\begin{figure}
  \includegraphics[width=\columnwidth]{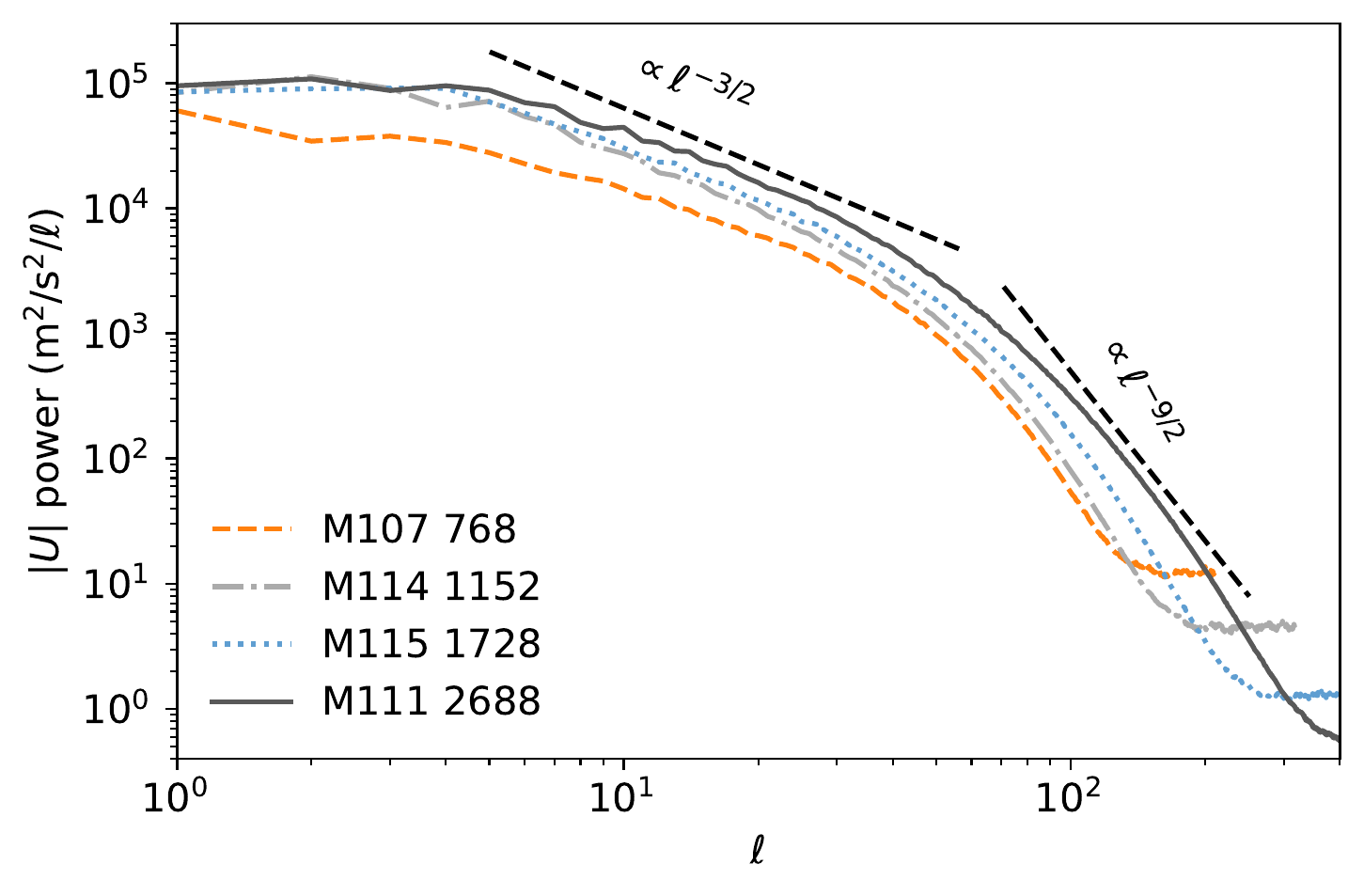}
  \includegraphics[width=\columnwidth]{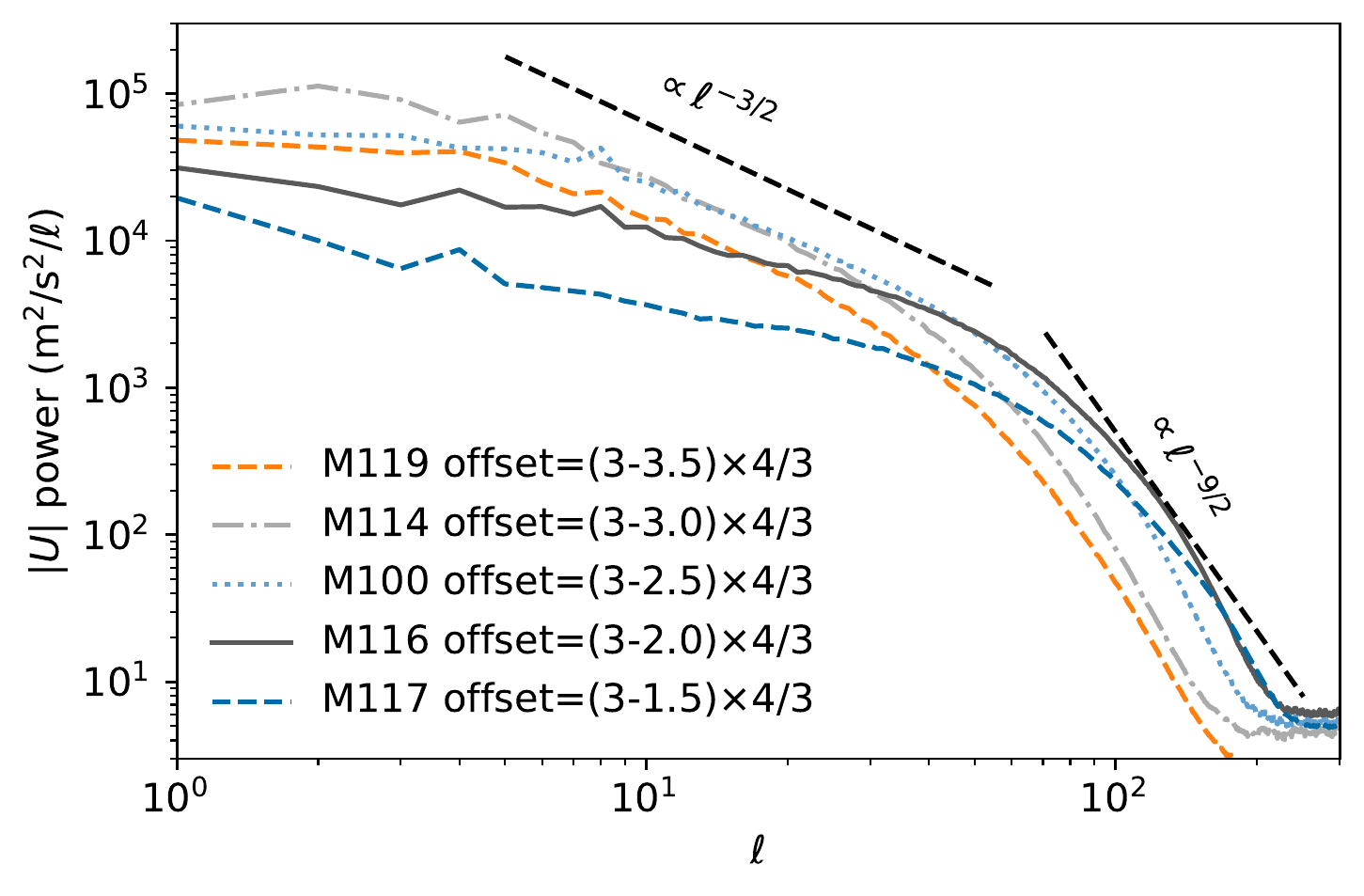}
  \caption{Spatial power spectra of velocity magnitude at radius \unit{1900}{\Mm}. Each line is the average of hundreds of dumps to remove short-term spectral variations. \textbf{Top}: Simulations with $1000\times$ ($10^3$) heating factor for different grid resolutions as indicated in the legend. \textbf{Bottom}: Spectra for different heating factors from $1152^3$-grid simulations (table 1, \papI) as indicated in the legend. The heating factor for each case is given as the log heating rate subtracted from $3$ as given in the bracket in the legend. For example, the heating factor simulation M119 is $10^{3.5}$. Each heating run has been shifted according to the scaling $u_\mathrm{IGW} \propto L^{2/3}$, see section 5.2 in \papI. Power laws have been added with dashed lines to guide the eye.  Simulations shown that are not included in \Tab{tab:run_table} are discussed in detail in \papI.}
  \lFig{convergence-spec}
\end{figure} 
\begin{figure}
  \includegraphics[width=\columnwidth]{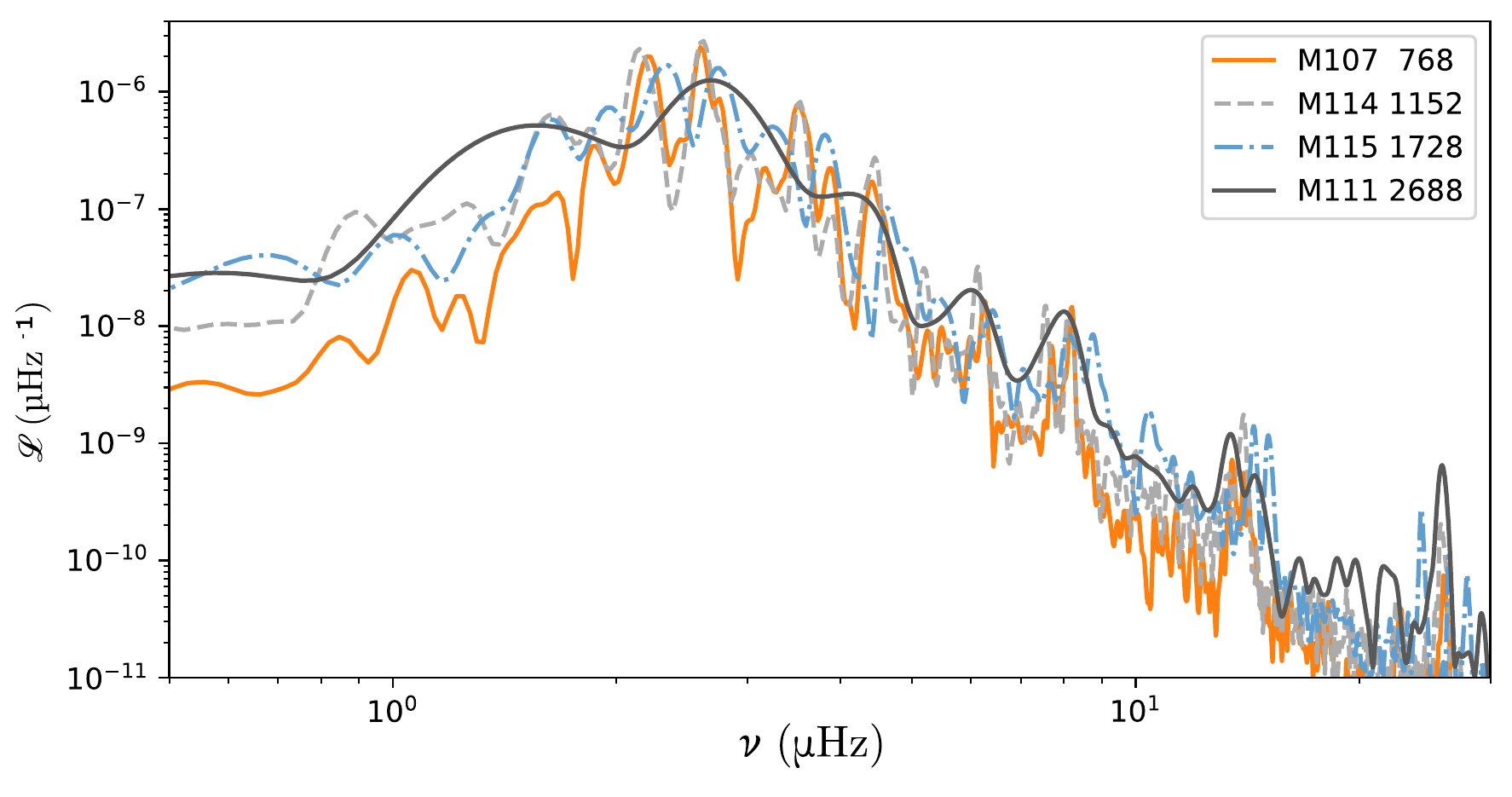}
  \includegraphics[width=\columnwidth]{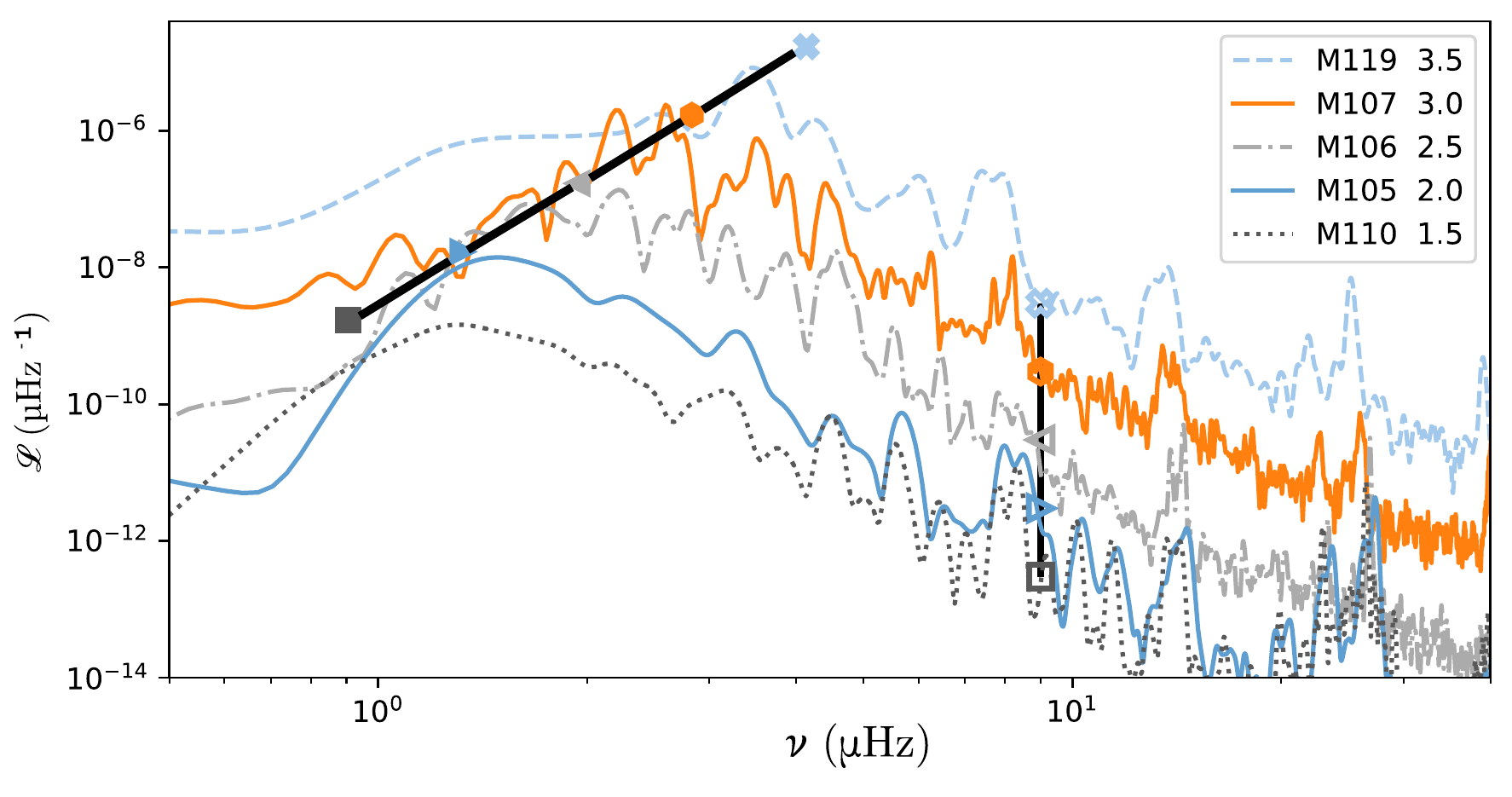}
  \caption{
    Temporal power spectral density of the unity-subtracted, relative, hemispherically integated luminosities $\mathcal{L}$, averaged over all eight lines-of-sight (\Sect{astero-analysis}), and using for each simulation the time series shown in \FigTwo{Resolution_lums_time_series}{Heating_lums_time_series} at radius \unit{1900}{\Mm}. Spectra have different frequency resolutions due to the different simulated time range available and/or adopted in the analysis.
    Top: Simulations with $1000\times$ heating factor for different grid resolutions as indicated in the legend.
    Bottom: Spectra for different log heating factors based on $768^3$-grid simulations, except M119 which is $1152^3$ grid. Filled markers are added at frequencies $\propto L^{1/3}$ reflecting the scaling of the convective frequency. 
    The vertical position of the markers is $\propto L^2$ and the offset is chosen to approximately match the power of simulation M107. A second set of open markers connected with a vertical line is added at \unit{9}{\microhz} with the same heating factor scaling.  Simulations shown that are not included in \Tab{tab:run_table} are discussed in detail in \papI.}
    \lFig{temporal_power_spectrum}
\end{figure}
\begin{figure}
  \includegraphics[width=\columnwidth]{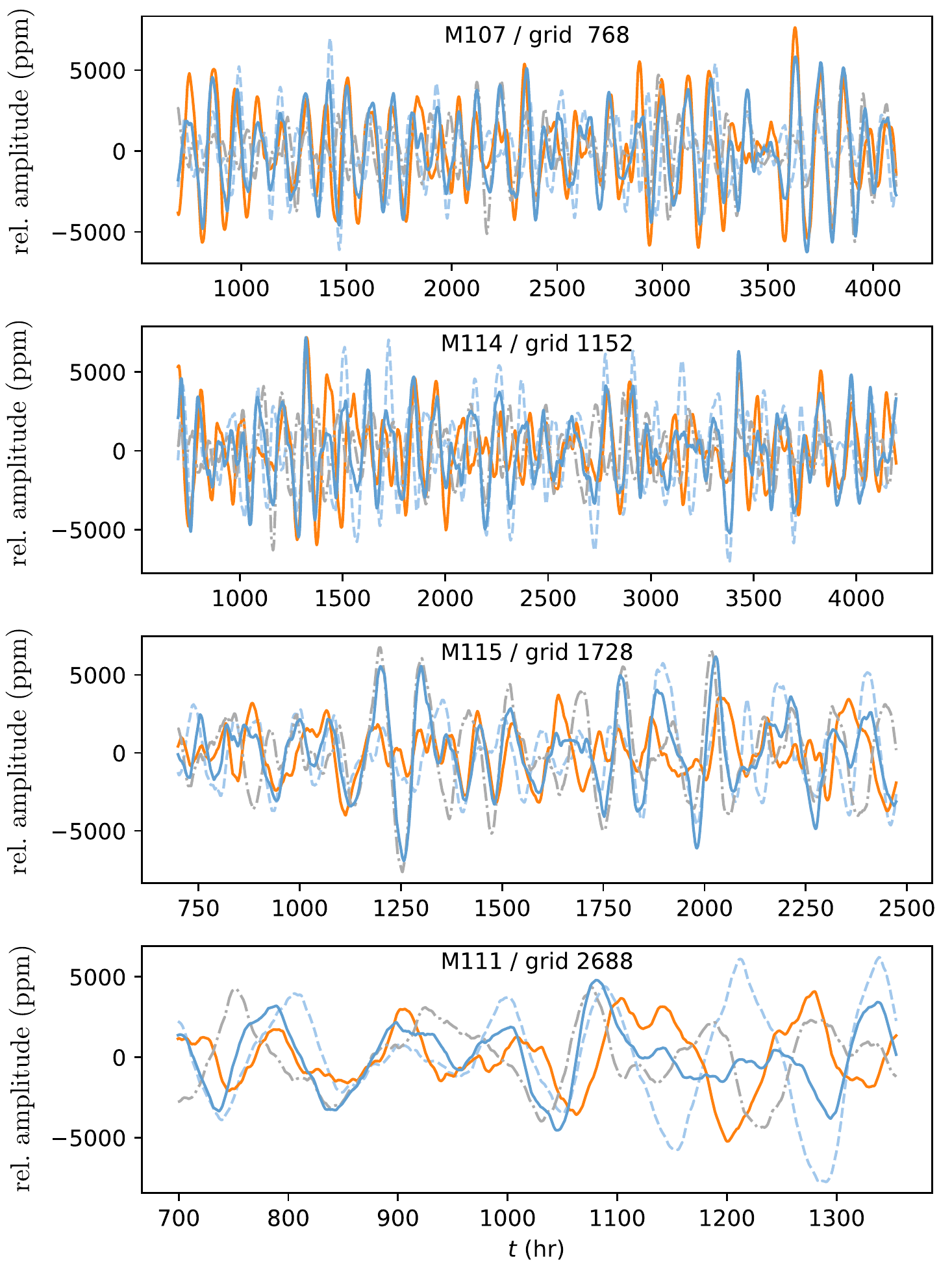}
  \caption{Time series of first four line-of-sight hemispherically integrated luminosities at radius  \unit{1900}{\Mm} for the resolution series used for the temporal spectra shown in \Fig{temporal_power_spectrum}. The text label gives the simulation name and the grid size.}
  \lFig{Resolution_lums_time_series}
\end{figure}
\begin{figure}
  \includegraphics[width=\columnwidth]{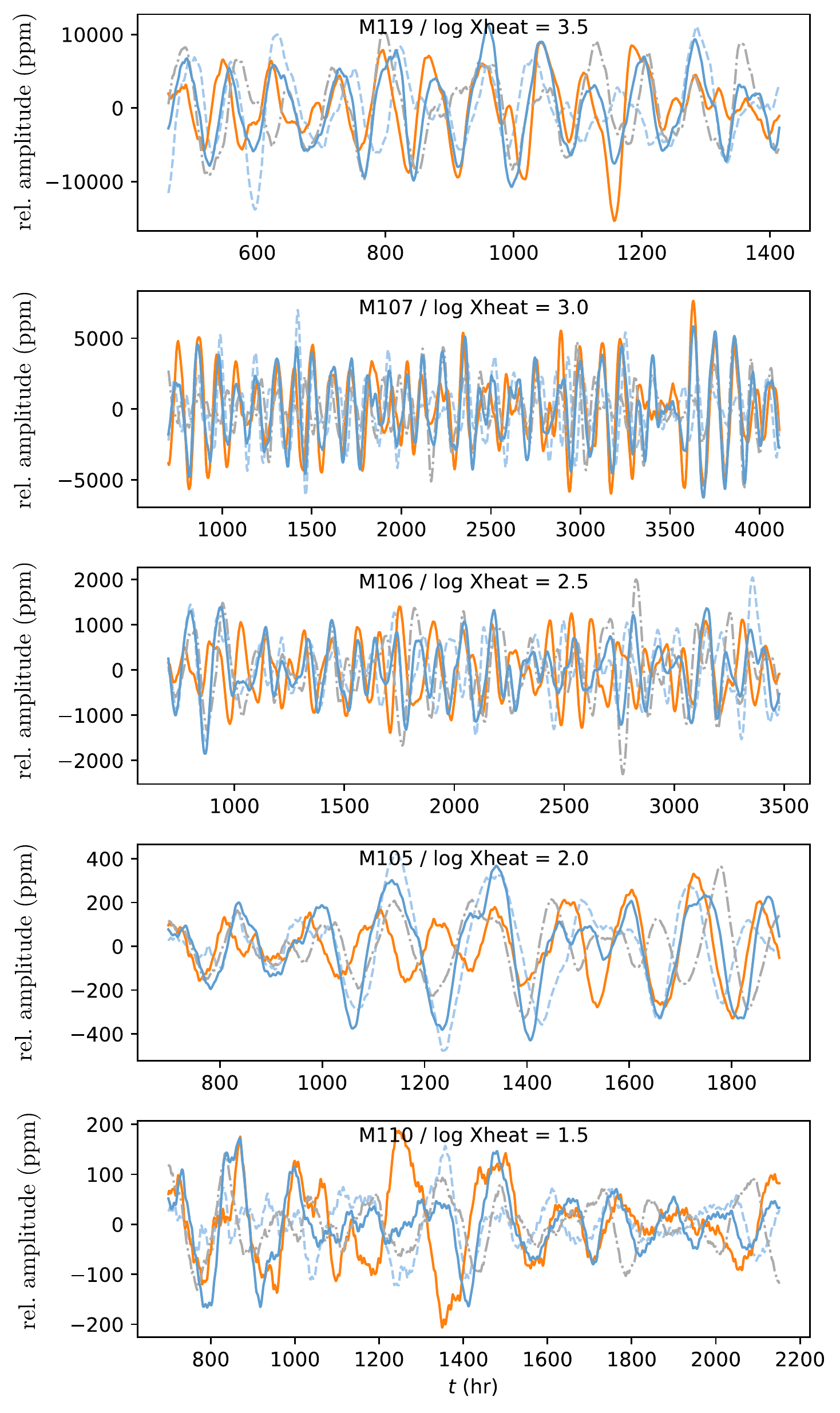}
  \caption{Time series of first for line-of-sight hemispherically integrated luminosities at radius  \unit{1900}{\Mm} used for the heating series temporal spectra shown in \Fig{temporal_power_spectrum}. The text label gives the simulation name and the log of the heating factor.}
  \lFig{Heating_lums_time_series}
\end{figure}

The dependence of IGW spectra driving luminosity deserves attention as well since all of our simulations are at a considerable boost factor. Demonstrating that spectra are self-similar not only under grid refinement but also as a function of heating factor is key for extrapolating from boosted simulations to nominal heating. We have already shown in \papI\ (section 5.2) that for heating rate factors $\leq 100$ for an $1152^3$ grid and $\leq316$ for a $752^3$ flow velocities start to depart from scaling relations, thereby indicating decreasing levels of accuracy. 
\FigTwo{convergence-spec}{temporal_power_spectrum} show both spatial and temporal spectra as a function of heating factor. The two lowest heating-factor simulations show somewhat flatter spatial spectra that extend relatively to higher anglar degree $l$. The three higher heating rate simulations agree well with each other when scaled according to $U_\mathrm{IGW} \propto L^{2/3}$.

The temporal spectra are shown in the bottom panel of \Fig{temporal_power_spectrum}. The frequency-integrated power spectrum density $\int \mathcal{L} d \nu$ scales linearly with the heating factor, with a deviation from the linear fit of $-2.5$, $-6.8$, $7.5$, $20.4$ and $-14.2\%$ from low to high heating factor. Markers are positioned at the convective frequency corresponding to the respective heating factor with vertical shifts reflecting the just mentioned scaling. The peaks of the spectra evolve with heating factor with almost the same rate as the expected convective frequency. Such a correlation was recently reported by \citet{LeSaux:22}. The same analysis at a larger radius of \unit{2400}{\Mm} showed an even smaller relationship between spectrum peak frequency and heating factor.

A second set of open markers at \unit{9}{\microhz} reflects the scaling found for the integrated power spectral density, and indicates that this scaling applies not only approximately to the peak power but also at other frequencies. Overall it appears that the spectra are mostly self-similar with respect to heating factor.

\section{Discussion}\lSect{discussion}
Based on 3D hydrodynamic simulations 
of a \unit{25}{\Msun} star near the beginning of its main sequence evolution we have investigated the excitation and spectrum of IGWs in the stable layer excited by core convection. The key convection and convective boundary properties of these simulations have been reported in \papI. The key results of this paper can be summarized as follows:
\begin{itemize}
  \item Deep inside the convective core the flow is well represented by a fully-developed Kolmogorov power spectrum. However, near the convective boundary the flow is dominated by horizontal flow topologies, in particular the boundary-separation flow wedges (cf. \papI) where unstable flow morphologies with a wide range of scales are generated. 
  \item The broad range of scales of fluid motions near the boundary excites IGWs with a broad range of spherical harmonic degrees $l$ and radial order $n$. The power remains significant up to $l \approx 30$ and the dispersion relations for radial orders up to $n \approx -10$ can be clearly distinguished in the \komega. These agree with the \GYRE\ predictions for the spherically averaged radial profile of the 3D simulation demonstrating that the oscillations are indeed IGWs. 
  \item The $n=-1$ g modes populate the convective boundary region where due to the choice of our initial profile from a \mesa\ stellar evolution model the profile of $N^2$ has a sharp peak (see \papI\ for details).
  \item The power in different IGWs fluctuates stochastically. This may reduce the cancellation effect when integrating photometric variability due to IGWs over a hemisphere. 
  \item IGWs have a broad distribution of power at moderate to high $l$ which corresponds, according to the dispersion relation of IGWs as seen in \Fig{k_omega_all_envelope}, to relatively high frequencies near the \brunt\ frequency. This power at high frequencies is attenuated by the cancellation in spherical integration, but not to the level of insignificance. 
  \item The combination of these two effects---the excitation of high-$l$ IGWs and reduced attenuation of incoherent IGWs---leads to a spectrum (right panels in \Fig{lums} and \emph{Original} in \Fig{lum_power_spectra_multi}) in which the power peaks in the frequency range that coincides with the convective frequency and then gradually tapers off toward the \brunt\ frequency.  
  \item When further applying the same pre-whitening procedure as is applied to observational data \citep{Bowman:2019ka,Bowman2020b} the low-frequency peaks that coincide with the convective frequency are removed resulting in a flat red noise spectrum shown in \FigTwo{lum_power_spectra_multi}{lum_power_spectrum}. 
  \item Spatial and temporal spectra show little dependence on grid resolution. The temporal spectra show a clear trend of the peak frequency with heating factor, with almost the same power as the convective frequency. The overall shape of the spatial and temporal spectra is self-similar with respect to heating factor.
  \item Since the convective velocity scales with $\propto L^{1/3}$ the convective frequency at nominal heating would be 10 times lower (thus $\approx \unit{0.27}{\mu\Hz}$) compared to our $1000\times$ simulation, and therefore challenging to access with current observations.
\end{itemize}

Our simulations do not include the effect of radiative damping (\Sect{stoch-excit}). Similar simulations to the ones analysed here have been presented in \papIII, where it has been shown that adding radiative diffusion does not alter the spectra enough to change the key findings of this work. Radiation diffusion is expected to happen predominantly near the surface which we do not include, and it may also  damp low temporal frequencies more than higher ones \citep{zahn:97}. How this damping manifests itself in the context of the other processes discussed here (e.g.\ \Sect{stoch-excit}) that lead to the formation of the IGW spectrum remains to be investigated. However, a key question is what role the IGWs excited by core convection play in generating the observed low-frequency excess in O and B stars.  \cite{lecoanetLowfrequencyVariabilityMassive2019} argue that the observed low-frequency excess cannot be due to IGW excitation from core convection, based on mainly two arguments relating to waves with $l \leq 3$. One key argument is that these low-$l$ waves should leave distinguishable features in the spectrum. Our simulations show that such individual features can be swamped in the forest of lines from the power residing at  higher $l$ ($15 \lessapprox l \lessapprox 45$). \cite{lecoanetLowfrequencyVariabilityMassive2019} discard such high-$l$ waves in their analysis presumably because they correspond to frequencies much higher than the convective frequencies, and it is maybe assumed that such high $l$ and high $\nu$ modes cannot be excited. Our simulations show that such waves are efficiently excited in boundary-flow features such as the boundary-layer separation wedges. Omitting high-$l$ waves would also be justified due to the expected cancellation effect in hemispheric integration which our simulations show to not be completely efficient, likely due their stochastic nature. In our simulated, hemispherically integrated spectrum the Lorentzian downturn expressed in the parameter $\gamma$ represents indeed the high-$l$ wave power attenuated by incoherent IGW cancellation. 

Although our simulations do not support the reasons for the suggestion of \cite{lecoanetLowfrequencyVariabilityMassive2019} that the observed red noise in massive stars is not caused by convective core excitation, we do agree with the notion that IGWs excited by convective core convection alone may not fully account for the observed spectra, based on conventional massive star stellar structure such as those used as initial model for our simulations. In order to obtain a low-frequency excess in a hemispherically-integrated mock observation spectrum through stochastic attenuation (\Sect{sec:attenuation}) the underlying eigenfrequencies need to extend to higher frequencies than in our simulations, because the IGW frequencies are limited by the \brunt\ frequency.

It has been suggested that the FeCZ zone would excite waves in the surface region \citep{cantiello:09,lecoanet:19,schultz:22}. Our simulations do not provide arguments that would speak against that. We speculate that the observed low-frequency excess is excited by a combination of core convection and near-surface convection zones, where the former contribute power to the lower frequencies and the surface convection adds power to the higher-frequency range. We expect that the reduced attenuation of incoherent IGWs would also apply to hemispheric integration of high-$l$ modes excited by near-surface convection modes. Indeed, \cite{Bowman2020b} observe a trend in the inferred morphology of the low-frequency power excess between younger and older stars. Such a transition also argues for both convective regions contribute to the excitation of IGWs observable at the surface, with a relative contribution of the two mechanisms that depends on the structure (i.e.\ age) of the star \citep{Bowman2022e}.

If the low frequency excess observed at the surface does indeed
originate from IGWs in the envelope, then the shape of the low-frequency excess could 
provide direct measurements of certain properties of the core and envelope. For instance,
we note that the location where the tail of the low-frequency excess falls to a negligible level coincides with the peak \brunt\ frequency of the envelope. Assuming that observational noise does not
overwhelm this signal, it could be used as a direct measurement of the properties of a star's envelope.
In a similar vein, we showed that certain coherent modes above the IGW background  
are variable. Individual modes can be seen in a spectrogram  (\Fig{sliding_fft}) to vary in
power during the simulation, and in some cases disappear before being stochastically re-excited.
If this is detected in observations, properties like the mean lifetime of these modes or time scale
after which they are re-excited could probe the core convection. Indeed, the sliding Fourier transforms of asteroseismic observations of O stars shown in \cite{Bowman:2019ib} reveal mode fluctuations, possibly on time scales of several dozen days which could correspond to this main fluctuation time scale of the IGW mode excitation. Recall that in a real star this time scale would be ten times longer than in our $1000\times$ heating simulations. 

We also identified a characteristic convective frequency
based on how long it takes on average for material in a convective wedge to 
travel from one end of the dipole until it separates from the boundary. This
excites a strong oscillations that peaks on the order of a few $\microhz$ (and again, ten times less at nominal heating). Detection of this
peak in observations could constrain the bulk speed of material traveling just under
the convective boundary.

Finally, the effect that adding radiation to our simulations will have
on the IGW spectrum will be considered in \papIII. 
This is expected to provide a damping mechanism that may affect the
overall distribution of power in the envelope, and may affect the 
time variability of discrete modes identified in a spectrogram. We also leave
a full parameter study on the effects of heating rate and rotation for a future work.

\section*{Acknowledgements}
WT acknowledges the support of the Natural Sciences and Engineering Research
Council of Canada (NSERC), 466479467.
FH acknowledges funding through an NSERC Discovery Grant. PRW acknowledges funding through NSF grants 1814181 and 2032010. Both have been supported through NSF award PHY-1430152 (JINA Center for the Evolution of the Elements).
DMB gratefully acknowledges funding in the form of a senior postdoctoral fellowship from the Research Foundation Flanders (FWO; grant number: 1286521N), from UK Research and Innovation (UKRI) in the form of a Frontier Research grant under the UK government’s ERC Horizon Europe funding guarantee (SYMPHONY; PI Bowman; grant number: EP/Y031059/1), and a Royal Society University Research Fellowship (URF; PI Bowman; grant number: URF{\textbackslash}R1{\textbackslash}231631).
SB acknowledges NSERC funding through a Banting Fellowship. The
simulations for this work was carried out on the Compute Canada
Niagara supercomputer operated by SciNet at the University of Toronto,
and on the NSF Frontera supercomputer operated by TACC at the
University of Austin, Texas. The data analysis was carried on the
Astrohub online virtual research environment
(\url{https://astrohub.uvic.ca} and \url{https://www.ppmstar.org}) developed and operated by the Computational
Stellar Astrophysics group (\url{http://csa.phys.uvic.ca}) at the University
of Victoria and hosted on The Alliance Arbutus Cloud at the
University of Victoria.  This work has benefited from scientific interactions at the KITP program "Probes of Transport in Stars" in November 2021, and therefore supported in part by the National Science Foundation (NSF) under Grant Number NSF PHY-1748958.

  We thank
Conny Aerts, Mathias Michielsen, and May Gade Pedersen for very
inspiring discussions.

\section*{Data availability}
3D and spherically averaged 1D simulation outputs are available at
\url{https://www.ppmstar.org} along with python notebooks that have
been used to create plots in this paper.

\bibliographystyle{mnras}
\bibliography{zotero,hydro} 




\bsp	
\label{lastpage}
\end{document}

%% file: paper_resources/macros/formatting.tex
\newcommand{\lSect}[1]{{\label{sec:#1}}}
\newcommand{\lFig}[1]{{\label{fig:#1}}}

\newcommand{\lTab}[1]{{\label{tab:#1}}}

\newcommand{\Tabff}[1]{{\ref{tab:#1}}}
\newcommand{\Tab}[1]{{Table~\Tabff{#1}}}

\newcommand{\pan}[1]{{\textit{#1}}}

\newcommand{\FIGFF}[2]{{\ref{fig:#2}\pan{#1}}}

\newcommand{\FIG}[2]{{Fig.~\FIGFF{#1}{#2}}}
\newcommand{\Fig}[1]{{\FIG{}{#1}}}
\newcommand{\FigTwo}[2]{{\FIGS{}{#1} and \FIGFF{}{#2}}}
\newcommand{\FIGS}[2]{{Figs.~\FIGFF{#1}{#2}}}

\newcommand{\Sectff}[1]{{\ref{sec:#1}}}
\newcommand{\Sect}[1]{{\S\Sectff{#1}}}

%% file: paper_resources/macros/units.tex
\newcommand{\numberspace}{\ensuremath{\;}}

\newcommand{\unitstyle}[1]{\ensuremath{\mathrm{#1}}}
\newcommand{\hour}{\unitstyle{h}}

\newcommand{\natlog}[2]{\ensuremath{#1\times 10^{#2}}} 

\newcommand{\kilo}{\unitstyle{k}}
\newcommand{\Mega}{\unitstyle{M}}

\newcommand{\meter}{\unitstyle{m}}

\newcommand{\second}{\unitstyle{s}}

\newcommand{\Msun}{\ensuremath{\unitstyle{M}_\odot}}

\newcommand{\minute}{\unitstyle{min}} 
\newcommand{\yr}{\unitstyle{yr}}        
\newcommand{\days}{\unitstyle{d}}        
\newcommand{\km}{\kilo\meter}   
\newcommand{\Mm}{\Mega\meter}   
\newcommand{\Hz}{\unitstyle{Hz}}        

\newcommand{\Hpzero}{\ensuremath{\unitstyle{H_{P0}}}}

\newcommand{\unit}[2]{\ensuremath{#1\numberspace\mathrm{#2}}}